\documentclass[epj]{svjour}
\usepackage{graphics}
\usepackage{amsmath}
\usepackage{amssymb}
\newcommand{\uvec}[1]{\hat{\boldsymbol{#1}}}
\newcommand{\tsi}{\textnormal{ts}_i}
\newcommand{\tsav}{\langle \textnormal{ts} \rangle}

\begin{document}
\title{Towards extracting cosmic magnetic field structures from cosmic-ray arrival directions}
\author{Marcus Wirtz \and Teresa Bister \and Martin Erdmann
}                     
%
%
\institute{RWTH Aachen University, III. Physikalisches Institut A, Otto-Blumenthal-Str., 52056 Aachen, Germany \and \email{marcus.wirtz@rwth-aachen.de}}
\date{Received: date / Revised version: date}
%
\abstract{
We present a novel method to search for structures of coherently aligned patterns in ultra-high energy cosmic-ray arrival directions simultaneously across the entire sky.
This method can be used to obtain information on the Galactic magnetic field, in particular the integrated component perpendicular to the line of sight, from cosmic-ray data only.
Using a likelihood-ratio approach, neighboring cosmic rays are related by rotatable, elliptically shaped density distributions and the significance of their alignment with respect to circular distributions is evaluated. 
In this way, a vector field tangential to the celestial sphere is fitted which approximates the local deflections in cosmic magnetic fields if significant deflection structures are detected.
The sensitivity of the method is evaluated on the basis of astrophysical simulations of the ultra-high energy cosmic-ray sky, where a discriminative power between isotropic and signal-induced scenarios is found.
\PACS{
      {96.50.S-95.85.Ry}{Cosmic rays - astronomical observations}   \and
      {96.50.S-98.70.Sa}{Cosmic rays - galactic and extragalactic}
     } 
} 
\maketitle
\section{Introduction}

\label{sec:introduction}

It is generally assumed that ultra-high energy cosmic rays (UHECRs) of extragalactic origin are deflected in the Galactic magnetic field (GMF). 
This assumption for deflections is, on one hand, based on astronomical measurements of Faraday rotation and synchrotron radiation, which indicate magnetic fields of micro-Gauss strengths~\cite{Han2017}. 
On the other hand, measurements of the atmospheric depth of cosmic rays can be explained by a composition of light to medium-heavy nuclei with charge numbers $Z\ge 1$ ~\cite{Aab2015,Aab2014a,Aab2017}. 
Together, these measurements predict deflections of the nuclei of several tens of degrees within the galaxy compared to their original extragalactic directions~\cite{Stanev1996,Harari2000,Harari2002,Golup2009,Giacinti2010,Golup2011,Giacinti2011}. 
In previous analyses that aimed to verify such deflections of cosmic rays, local regions of arrival were examined for energy ordering, but no scientific evidence for particle deflections was found for any region~\cite{Abreu2011,Aab2014b,Aab2020}.
Recently, we introduced a fit method that determines the most probable extragalactic source directions by inverting the deflections that are caused by a specific GMF model and fitting a particle charge for each cosmic-ray event~\cite{Erdmann2019}.
The several thousand free parameters are fitted using the backpropagation method developed for neural network training~\cite{Tensorflow2015}.

\begin{figure*}[ht]
\centering
\resizebox{0.8\textwidth}{!}{\includegraphics{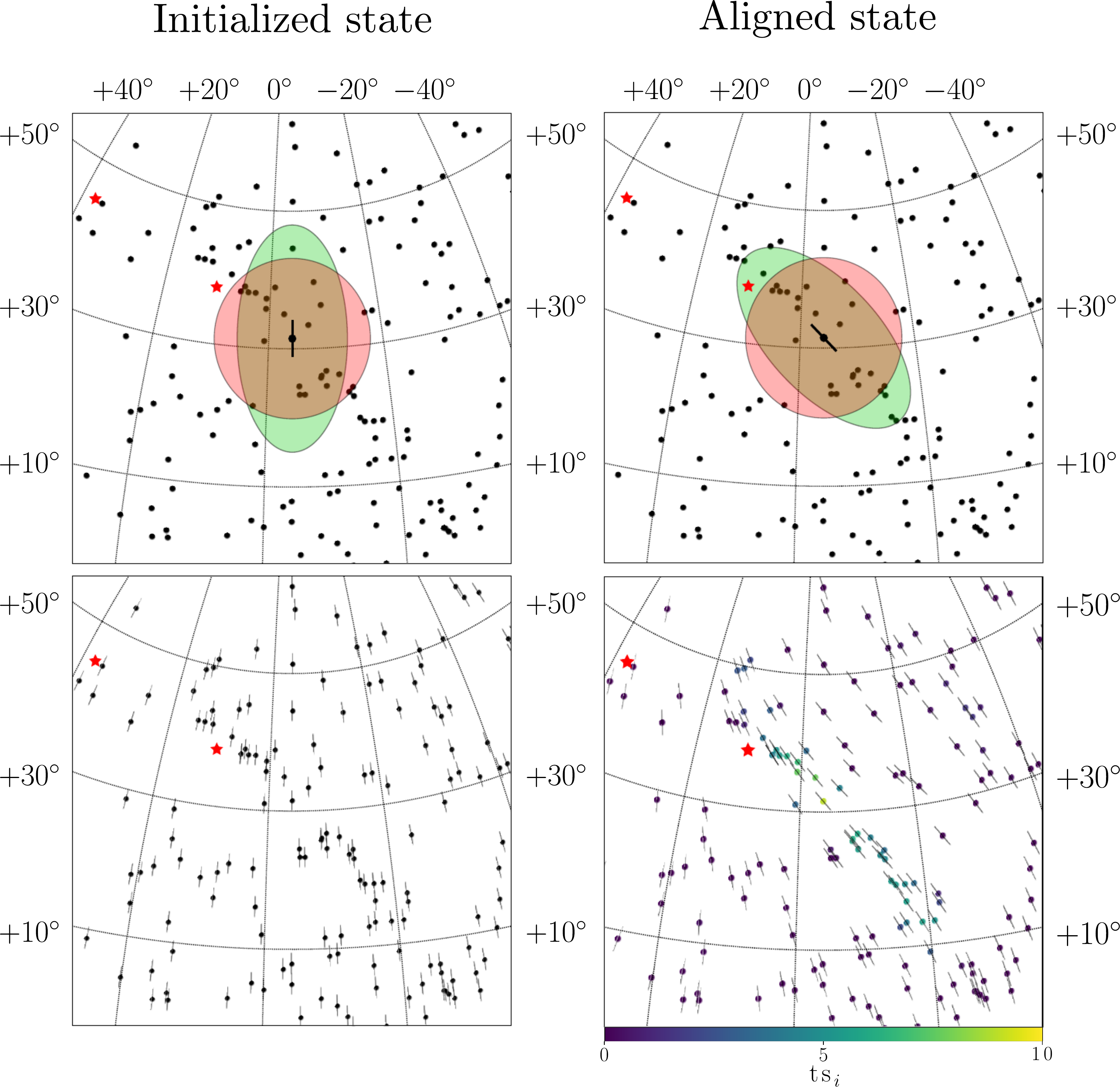}}
\caption{Concept of the COMPASS method demonstrated in an astrophysical simulation (cf. section~\ref{sec:compass_benchmark}). Circular symbols mark UHECR arrival directions, short lines denote the fitted orientation of coherent alignment, and the red stars show the directions of simulated sources. (\textit{Left}) State of the system at initialization and (\textit{right}) after the fit. (\textit{Top}) Shapes of the elliptical signal probability density function in green and the Gaussian background probability density function in red. (\textit{Bottom}) Local orientation of fitted alignment patterns. The color code in the lower right panel corresponds to the likelihood ratio resulting from the signal and background density functions (see text).}
\label{img:concept}
\end{figure*}

In this work, we present a novel approach in which all cosmic-ray arrival directions are simultaneously examined for alignment structures without relying on a certain GMF model~\cite{Wirtz2019b}. 
The method is independent of energy ordering and analyzes only the arrival directions above a minimum energy threshold. 
In order to quantify coherent directional deflections, elliptically shaped regions are employed whose orientation is optimized by the frequency of neighboring particles (cf. Fig.~\ref{img:concept}). 
Coherence of adjacent ellipses is realized by means of a spherical harmonic expansion which assigns the local orientations of the set of ellipses.

The method is formulated as a likelihood ratio where for each cosmic-ray arrival direction, it is checked whether the cosmic ray is part of a deflection pattern or rather a particle of isotropic arrival directions. 
As a null hypothesis, circular regions are used instead of elliptical regions to distinguish the effects of coherent deflections from overdensities. 
The likelihood ratio is employed as the objective function for adjusting the spherical harmonic functions that specify magnetic field deflections. 
Thus, the test statistics of all measured particles are used to answer the question whether coherent deflections exist in the cosmic-ray arrival directions.
If the answer to this question is positive, the orientations of the ellipses indicate the directional deflections caused by the GMF.
This novel approach is hereafter referred to as: COherent Magnetic Pattern Alignment in a Structure Search (COMPASS).

The work is structured as follows: 
First, the analysis strategy is presented, covering the tangent vector field, the definition of the likelihood ratio, and its normalization.
Two benchmark simulations are then introduced: one features simplified patterns of point sources to demonstrate the proof of concept and the other one is an advanced astrophysical simulation where UHECR nuclei from uniformly distributed sources are attenuated during propagation in the extragalactic universe.
The ability to reconstruct the coherent directional deflections of the GMF and the advantages of using a circular reference model in the likelihood ratio are demonstrated in the following two chapters.
Finally, the sensitivity of the method is investigated for both simulations, the simplified patterns, and the astrophysical universe.

\section{Analysis strategy}
\label{sec:strategy}

The objective of the COMPASS method is to find alignment patterns in UHECR arrival directions simultaneously across the entire sky.
In this approach, an adjustable vector field $\uvec{u}(\vartheta, \varphi)$ tangential to the local celestial sphere determines the orientation of elliptically shaped probability density functions (PDFs) that are centered on each cosmic-ray arrival direction.
Here, $\vartheta$ and $\varphi$ denote the polar angle and azimuthal angle, respectively, in a spherical coordinate system.
The likelihood that the distribution of neighboring arrival directions is better described by an elliptical PDF than by a background hypothesis is then evaluated to optimize the orientation of the ellipses' major axes for all cosmic-ray events in one single step.
In this way, the vector field $\uvec{u}(\vartheta, \varphi)$ locally aligns with elongated structures which are expected to occur from UHECR deflections in the GMF\footnote{~An intuitive analogy for the concept is the alignment of iron filings in magnetic fields.}.
Technically, a likelihood ratio (cf. equation~\eqref{eq:test_statistic}) serves as an objective function in a minimization based on gradient descent.
Additional constraints within the analysis can be accounted for by adding a corresponding penalization term to the objective function.

The basic concept of the COMPASS method is demonstrated in Fig.~\ref{img:concept} where the initial state of the system is shown in the left panel and the fitted state in the right panel.
Here, the initialization of the vector field $\uvec{u}(\vartheta, \varphi)$ is equal to the local unit vector $\uvec{e}_\vartheta$.
The upper panel shows the ellipsoidal PDF (green) and a corresponding Gaussian background PDF (red).
One can see that the orientation of the ellipse has changed after the fit where an alignment with a prominent pattern originating from the red marked source is clearly visible.
Additionally, the lower panel indicates the orientation of the adjusted vector field $\uvec{u}(\vartheta, \varphi)$ in the vicinity of the pattern.
The orientation has changed considerably only for the cosmic-ray events which are part of the pattern, whereas the ellipses of most of the isotropic events have not changed substantially during the fit.
This finding is also visualized by the color-coded likelihood ratio where high values are found only for events that are part of the pattern.

The method requires a high number of fit parameters, both for the parameterization of the vector field $\uvec{u}(\vartheta, \varphi)$ and the UHECR model for the likelihood ratio.
For the analyzed simulated data set of UHECRs with energies above $40$~EeV, the number of free fit parameters is of the order of $\mathcal{O}(1000)$.
Our method uses the software package {TensorFlow} \cite{Tensorflow2015} to perform a gradient descent-based optimization in this high dimensional parameter space.
To enable the computation of gradients within the scope of the backpropagation technique used in the field of machine learning, all operations of this analysis (cf. following subsections) --- spherical harmonics expansion, parameterization of density functions, vector algebra operations, likelihood ratio --- were written with the {TensorFlow} API.

\subsection{Tangent vector field}

A particular challenge is the parameterization of the tangent vector field $\uvec{u}(\vartheta, \varphi)$ which is meant to describe the orientation of deflection patterns caused by the GMF.
The large-scale component of the GMF is most likely responsible for patterns of coherent deflection \cite{Erdmann2016,Farrar2017}.
Thus, the orientation of patterns is expected to vary only slightly within a local domain of the sky.

Here, the adaptable vector field is first realized by a constant vector field $\uvec{u}_0(\vartheta, \varphi)$ which serves as an initialization and is then modified locally by an angle $\Psi(\vartheta, \varphi)$.
To preserve the local coherence of the resulting vector field $\uvec{u}(\vartheta, \varphi)$, the modification angle is parameterized by a spherical harmonics expansion of order $k$:
\begin{equation}
\label{eq:psi}
\Psi(\vartheta, \varphi) = \sum_{\ell=0}^{k} \sum_{m=-\ell}^{\ell} a_\ell^m Y_\ell^m(\vartheta, \varphi) \, ,
\end{equation}
where $Y_\ell^m(\vartheta, \varphi)$ are the spherical harmonics functions and $a_\ell^m$ represent a set of free fit parameters to model any continuous differentiable function on the surface of the sphere.
Rapid variations on small angular scales can be suppressed by demanding the order $k$ of the spherical harmonics expansion to have an upper limit of $k=5$.
Typical examples for this value $k$ used in the analysis are $k=4$ and $k=5$ which yield $25$ and $36$ free fit parameters, respectively.
The resulting modification $\Psi$ for a certain direction $\uvec{r}$ is then described by a rotation of $\uvec{u}_0$ around the axis $\uvec{r}$:
\begin{equation}
\label{eq:vector_field}
\uvec{u}(\vartheta, \varphi) = R (\uvec{r}(\vartheta, \varphi), \Psi(\vartheta, \varphi)) \, \uvec{u}_0 (\uvec{r}(\vartheta, \varphi)) \; ,
\end{equation}
where the two arguments of the rotation matrix $R$ are the rotation axis and angle, respectively.
Note that here the polar angle $\vartheta$ is defined as being consistent with the Galactic latitude $b$; thus, in Cartesian coordinates $\uvec{r}$ is given by:
\begin{equation}
\label{eq:spherical_coordinates}
\uvec{r}(\vartheta, \varphi) = (\cos \vartheta \, \cos \varphi, \, \cos \vartheta \, \sin \varphi, \, \sin \vartheta)^T
\end{equation}

For the initialized vector field $\uvec{u}_0 (\uvec{r})$, three approaches were investigated in this work.
The \textit{hairy ball theorem} states that there exists no nonvanishing continuous tangent vector field on the surface of the three-dimensional sphere \cite{Renteln2013}.
Thus, $\uvec{u}_0$ always exhibits at least one region on the sphere where the vector field either radially diverges at a certain point or where it circularly curls around it.
The following three initializations were used:
\begin{itemize}
  \item \textbf{JF12 GMF}: An intuitive approach is to initialize the fit with the best guess of the pattern orientations, e.g. the predictions of the currently most reliable GMF model, namely that developed by Jansson \& Farrar \cite{Jansson2012a} (JF12). 
  Here, to obtain the local direction of deflection in the direction of $\uvec{r}$, a magnetically highly rigid particle of $10^{20}$~eV is backtracked, leaving the Galaxy in direction $\uvec{r}'$. 
  Then, the local tangent vector field is defined as $\uvec{u}_0 (\uvec{r}) = \uvec{r} \times (\uvec{r} \times \uvec{r}') / C$ where $C$ is determined by the normalization according to $|| \uvec{u}_0 || = 1$.
  \item \textbf{Galactic meridians}: Here, the local tangent vector $\uvec{u}_0 (\uvec{r})$ is equal to the local spherical unit vector $\uvec{e}_\vartheta$ in the Galactic coordinate system. The advantage of this initialization is that it is independent of a certain GMF model, while an overall symmetry with respect to the Galactic plane is still maintained. Additionally, certain models favor a general deflection preference towards the Galactic plane~\cite{Farrar2015}, which is approximately realized in this case.
  \item \textbf{Equatorial meridians}: In analogy to the Galactic meridians, here the initialization is equal to the unit vector $\uvec{e}_\theta$ in the Equatorial coordinate system. This initialization has the advantage that one of the two points of divergence is located in the blind region of a ground-based Observatory. Here, it is used only as a crosscheck for the fit reliability.
\end{itemize}

\begin{figure}[h]

\centering
\resizebox{0.49\textwidth}{!}{\includegraphics{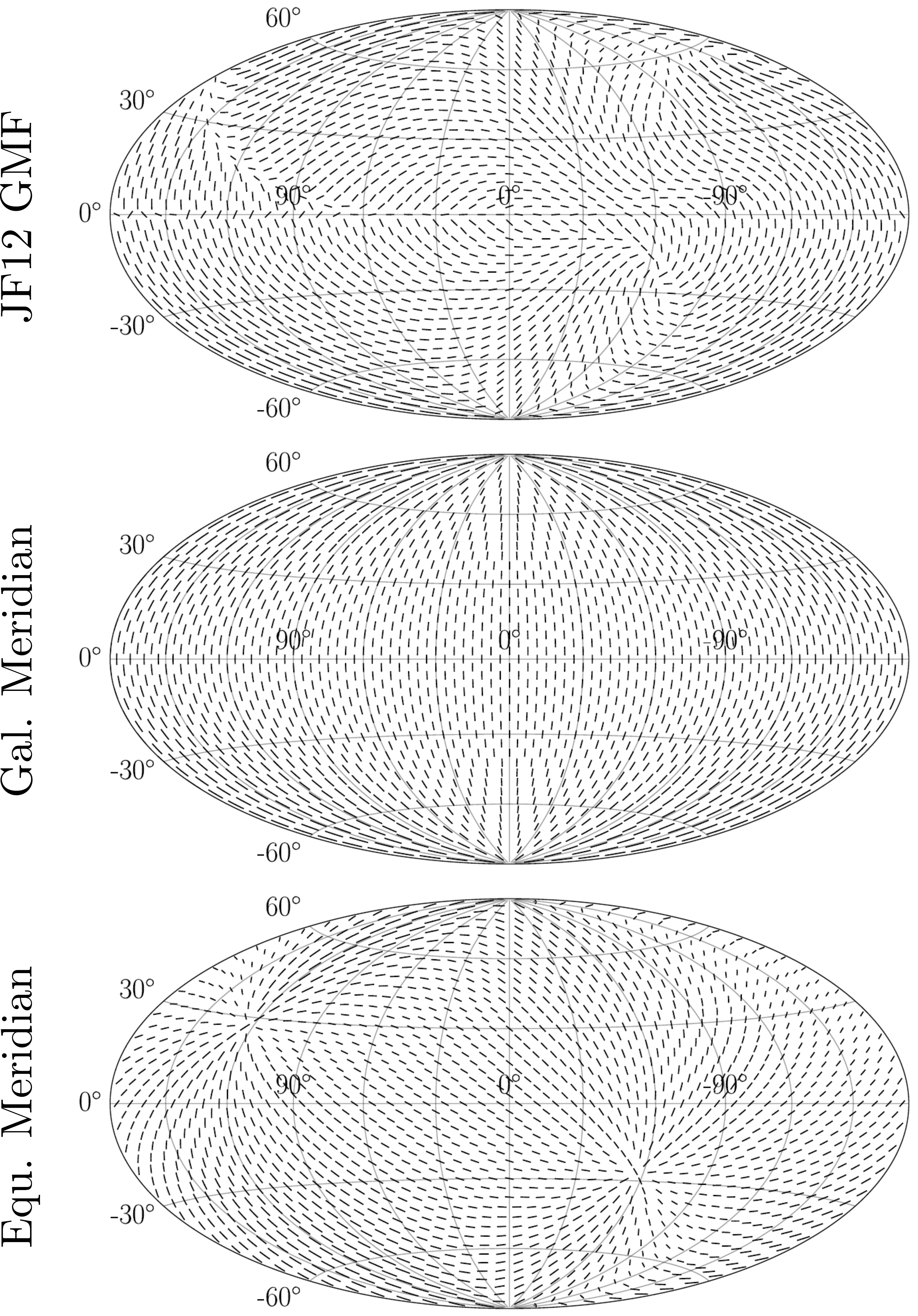}}
\caption{Visualization of the utilized tangent vector fields $\uvec{u}_0$ in the Galactic coordinate system. (\textit{Top}) Local orientation of deflection patterns from simulations in the JF12 field calculated by the displacement of an $R=10^{20}$~eV particle. (\textit{Middle}) Meridians in Galactic coordinates along the local $\uvec{e}_\vartheta$ unit vector. (\textit{Bottom}) Meridians in Equatorial coordinates.}
\label{img:vector_field}
\end{figure}

A visualization of the three tangent vector field initializations $\uvec{u}_0$ is presented in Fig.~\ref{img:vector_field}.
Regions with curls or divergences of the vector field can be seen in all three initializations.
At these locations, the analysis exhibits a decreased sensitivity to find locally aligned structures as the underlying vector field $\uvec{u}_0$ cannot describe them.
For the initialization of \textit{JF12 GMF}, two of these features are visible at Galactic coordinates $(l, b) \approx (90^\circ, 0^\circ)$ and $(l, b) \approx (-60^\circ, -20^\circ)$.
The \textit{Galactic} and \textit{equatorial meridians} initializations exhibit two divergences at the northern and southern poles of the respective coordinate system.
An example of the working principle of the modification function $\Psi(\vartheta, \varphi)$ for the \textit{Galactic meridians} initialization is shown in the lower panel of Fig.~\ref{img:concept}.

For the \textit{JF12 GMF} initialization --- depending on the reliability of the model predictions --- it may be beneficial for the sensitivity to include a penalization term for large model deviations in the objective function.
This penalization can be achieved by limiting the integrated squared amplitude of $\Psi$ over the entire celestial sphere as:

\begin{equation}
\label{eq:gmf_objective}
F = \frac{1}{4 \pi} \int_0^{2\pi} \int_{-\pi/2}^{\pi/2} 
  \Psi^2(\vartheta, \varphi) \, \cos(\vartheta) \, d\vartheta \, d\varphi \, .
\end{equation}

As a potential improvement in scenarios where many directions exhibit alignment patterns, the tangent vector field $\uvec{u}$ may be directly defined in the form of vector spherical harmonics (VSH) \cite{Barrera1985}.
In this way, positions of curls and divergences of the vector field can be shifted dynamically over the sky during the fit. 
Thus, they will likely stall in sky regions without noteworthy contributions to the likelihood ratio, i.e. in regions without prominent alignment patterns.

\subsection{Maximum likelihood ratio}

The COMPASS method evaluates the distribution of arrival directions around each cosmic-ray event $i$ in order to search for the existence of an elongated structure.
Here, the likelihood is defined in analogy to the approach in \cite{Aab2018a}:
the total UHECR sky model consists of a sum of a signal part $\mathcal{E} (\uvec{r}) \, \mathcal{S}_i (\uvec{r})$ (with the contribution $| f_i |$) which captures elongated patterns and a purely isotropic part $\mathcal{E} (\uvec{r})$ which represents the geometrical exposure of the observatory \cite{Sommers2000}.
This likelihood $\log (\mathcal{L}_i^\mathcal{S})$ is then compared to a suitable reference model by calculating the likelihood ratio.
The fundamental difference with respect to the approach in \cite{Aab2018a} is that each cosmic-ray event $i$ provides a separate density function $\mathcal{S}_i (\uvec{r})$ including a fit parameter $f_i$ which describes the contribution of the respective event to the log-likelihood as:

\begin{equation}
\label{eq:signal_likelihood}
\begin{aligned}
\log (\mathcal{L}_i^\mathcal{S})= \sum_j^{N_\textnormal{tot}} & \log \left[ \, | f_i | \times \mathcal{E} (\uvec{\Theta}_j) \, \mathcal{S}_i(\uvec{\Theta}_j) \right. \\ 
& \left. \qquad + (1 - | f_i |) \times \mathcal{E} (\uvec{\Theta}_j) \, \right] \, ,
\end{aligned}
\end{equation}

where $N_\textnormal{tot}$ is the total number of events in the data set and $\uvec{\Theta}_j$ the unit vector of the arrival direction of cosmic ray $j$.
Here, the signal and background contributions, $\mathcal{E} (\uvec{r}) \, \mathcal{S}_i (\uvec{r})$ and $\mathcal{E} (\uvec{r})$, are normalized over the surface $A$ of the sphere:

\begin{equation}
\label{eq:normalization_likelihood}
\iint_A \mathcal{E} (\uvec{r}) \, \textnormal{d}A = 1 \quad \textnormal{and} \quad \iint_A \mathcal{E} (\uvec{r}) \, \mathcal{S}_i(\uvec{r}) \, \textnormal{d}A = 1 \; .
\end{equation}

The set of $f_i$ represents a total number of $N_\textnormal{tot}$ free fit parameters which are initialized with a value close to zero.
Thus, the total number of free parameters of the COMPASS method is $n_\textnormal{fit} = (k+1)^2 + N_\textnormal{tot}$, where the $(k+1)^2$ part comes from the spherical harmonics coefficients $a_\ell^m$.

The signal part $\mathcal{S}_i (\uvec{r})$ of equation~\eqref{eq:signal_likelihood} is constructed as an elliptically shaped density function which is centered at cosmic-ray direction $\uvec{\Theta}_i$.
The major axis is aligned with the local direction of the tangent vector field $\uvec{u}_i \equiv \uvec{u}(\vartheta_i, \varphi_i)$.
Since the GMF is not well known, there is no accurate mathematical description for the expected shape and size of a deflection pattern.
Here, the density function $\mathcal{S}_i$ is parameterized on the basis of a non-symmetrical Gaussian distribution where the width follows an ellipse equation as

\begin{equation}
\label{eq:ellipse_equation}
\mathcal{S}_i (\uvec{r}) = C \times \exp \left(- \frac{(\uvec{r} \cdot \uvec{u}_i)^2}{\delta_{\text{max}}^2} -\frac{(\uvec{r} \cdot (\uvec{\Theta}_i \times \uvec{u}_i))^2}{\delta_{\text{min}}^2} \right) \; ,
\end{equation}

for all directions $\uvec{r}$ located in the same hemisphere as cosmic ray $i$, i.e. $\uvec{\Theta}_i \cdot \uvec{r} \geq 0$.
In the opposite hemisphere of the sky the density function is set to zero.
The hyperparameters $\delta_{\text{max}}$ and $\delta_{\text{min}}$ denote the angular extent in the direction of the ellipses' semi-major and semi-minor axes, respectively.
$C$ denotes a normalization factor which is investigated in section~\ref{sec:normalization}.

For every cosmic ray $i$, the minimal distance $\delta_{i, j}^{\perp}$ between the direction $\uvec{\Theta}_j$ of the neighboring cosmic ray $j$ and the orthodrome $\zeta$, as defined by $\uvec{\Theta}_i$ and $\uvec{u}_i$, is given by the relation:

\begin{equation}
\sin (\delta_{i, j}^{\perp}) = \frac{\uvec{\Theta}_j \cdot (\uvec{\Theta}_i \times \uvec{u}_i)}{|| \uvec{\Theta}_i \times \uvec{u}_i ||} \; .
\end{equation}

Since $\uvec{\Theta}_i$ and $\uvec{u}_i$ are unit vectors, the term $|| \uvec{\Theta}_i \times \uvec{u}_i ||$ is equal to one.
Thus, the numerator of the second term in the exponential function of equation~\eqref{eq:ellipse_equation} can be identified as the transverse displacement of cosmic-ray direction $\uvec{\Theta}_j$ relative to the fitted orientation $\uvec{u}_i$.
Likewise, the great-circle distance along the orthodrome $\zeta$ -- and therefore along the pattern orientation --- is given by $\sin (\delta_{i, j}^{\parallel}) = \uvec{\Theta}_j \cdot \uvec{u}_i$.
Thus, for small angles, i.e. $\sin \delta_{i, j} \approx \delta_{i, j}$, equation~\eqref{eq:ellipse_equation} can be identified as a two-dimensional Gaussian distribution on the sphere where contour lines of equal function values $\mathcal{S}_i$ follow an ellipse equation:

\begin{equation}
\left( \frac{\delta_{i, j}^{\parallel}}{\delta_{\text{max}}} \right)^2 
+ \left( \frac{\delta_{i, j}^{\perp}}{\delta_{\text{min}}} \right)^2 = \textnormal{const} \; .
\end{equation}

\begin{figure}
\centering
\resizebox{0.49\textwidth}{!}{\includegraphics{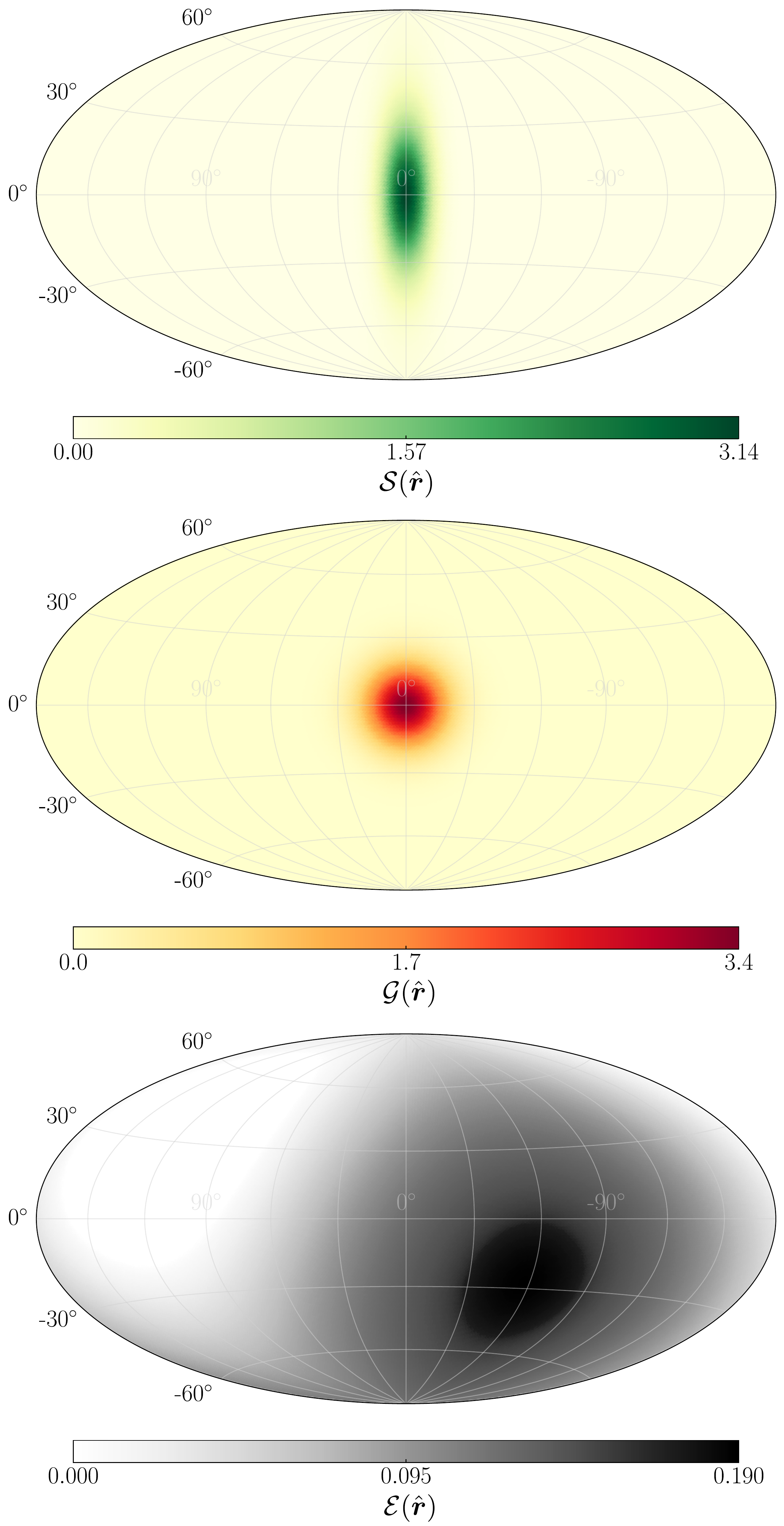}}
\caption{Normalized probability density distributions for (\textit{top}) signal part $\mathcal{S}_i (\uvec{r})$, (\textit{middle}) the Gaussian reference part $\mathcal{G}_i (\uvec{r})$, and (\textit{bottom}) the geometrical exposure $\mathcal{E} (\uvec{r})$. For the signal and Gaussian reference parts, the cosmic-ray direction $\uvec{\Theta}_i$ is centered at Galactic coordinates $\vartheta = 0^\circ$ and $\varphi = 0^\circ$ and the ellipse geometry is $\delta_\textnormal{max}=30^\circ$ and $\delta_\textnormal{min}=10^\circ$ where the major axis is aligned with the $\uvec{e}_\vartheta$ unit vector. The geometrical exposure is given by the geometry of the Pierre Auger Observatory with a maximum zenith angle of $80^\circ$.}
\label{img:density_distributions}
\end{figure}

For the reference model of the likelihood ratio, two approaches were explored in this work: a purely isotropic hypothesis following the geometrical exposure $\mathcal{E} (\uvec{r})$ and a Gaussian reference model with identical signal strength $f_i$ (compared to the elliptical ones) to cancel out overdensities.

\begin{itemize}
  \item \textbf{Isotropic reference model} $\mathcal{E} (\uvec{r})$: The highest sensitivity to reject an isotropic scenario is obtained by testing explicitly against this hypothesis in the likelihood ratio. Thus, each cosmic ray $i$ provides the same log-likelihood contribution:
  \begin{equation}
  \label{eq:reference_likelihood_isotropy}
\log (\mathcal{L}_i^\mathcal{R})= \sum_j^{N_\textnormal{tot}} \log ( \mathcal{E} (\uvec{\Theta}_j) )  \, ,
  \end{equation}
  \item \textbf{Gaussian reference model} $\mathcal{G}_i (\uvec{r})$: Here, the reference model is provided by equation~\eqref{eq:ellipse_equation} where both the major axis and minor axis radii are set to $\sqrt{\delta_\textnormal{max} \times \delta_\textnormal{min}}$ with respect to the ellipse dimensions of the signal distribution. By choosing the geometric average of both dimensions, the effective solid angle of the symmetric reference model is unchanged and equation~\eqref{eq:ellipse_equation} can be written as a symmetric Gaussian-like distribution:
  \begin{equation}
  \mathcal{G}_i (\uvec{r}) = C \times \exp \left(- \frac{ \sin^2( \, \sphericalangle ( \uvec{r} , \uvec{\Theta}_i ) \, )}{\delta_{\text{max}} \times \delta_{\text{min}}} \right) \; .
  \end{equation}
  In the log-likelihood ratio, the same value for the contribution $f_i$ as in equation~\eqref{eq:signal_likelihood} is chosen to evaluate solely the difference between an elliptically shaped and a symmetrical pattern:
\begin{equation}
\label{eq:reference_likelihood_gaussian}
\begin{aligned}
\log (\mathcal{L}_i^\mathcal{R})= \sum_j^{N_\textnormal{tot}} & \log \left[ \, | f_i | \times \mathcal{E} (\uvec{\Theta}_j) \, \mathcal{G}_i(\uvec{\Theta}_j) \right. \\ & \left. \qquad + (1 - | f_i |) \times \mathcal{E} (\uvec{\Theta}_j) \, \right] \, ,
\end{aligned}
\end{equation}

  During the TensorFlow fit, the gradient of $f_i$ is computed only with respect to the signal contribution in equation~\eqref{eq:signal_likelihood} in order to prevent active adaption of the reference model $\mathcal{R}$.
\end{itemize}

Examples of the normalized probability density function of the elliptically shaped signal part $\mathcal{S} (\uvec{r})$, the symmetric reference part $\mathcal{G} (\uvec{r})$, and the geometrical exposure $\mathcal{E} (\uvec{r})$ are visualized in Fig.~\ref{img:density_distributions}.

For the objective function of the fit, each cosmic ray contributes with a separate log-likelihood ratio $\tsi$ as:

\begin{equation}
\label{eq:individual_test_statistic}
\tsi = 2 \times [ \, \log (\mathcal{L}_i^\mathcal{S}) - \log (\mathcal{L}_i^\mathcal{R})  \, ] \; .
\end{equation}

For an isotropic arrival distribution, each individual $\tsi$ follows a $\chi^2$ distribution with the degrees of freedom corresponding to the free fit parameters according to Wilks' theorem \cite{Wilks1938}.
Therefore, the corresponding sum of all individual test-statistic contributions is Gaussian distributed according to the central limit theorem.
The latter statement holds only if the individual contributions are independent of each other, which is not entirely the case in this application since the density functions of neighboring cosmic rays overlap.
Nevertheless, it has been explicitly checked in Monte Carlo simulations of isotropic arrival distributions that the summed test statistics $\sum_i \tsi$ approximately follows a Gaussian distribution.
Thus, the average test statistic $\tsav$ provides a well-defined metric to evaluate the pattern alignment over the entire sphere:
\begin{equation}
\label{eq:test_statistic}
 \tsav = \frac{1}{N} \sum_i^{\text{N}} \tsi \; .
\end{equation}
To maximize equation~\eqref{eq:test_statistic}, the negative average test statistic is chosen as the objective function for the gradient descent. 
Additionally, in the case of the \textit{JF12 GMF} initialization of the tangent vector field, the objective term $F$ from equation~\eqref{eq:gmf_objective} is added.  
Thus, the total objective function $J$ exhibits one hyperparameter $\lambda_F$ which represents the confidence in the GMF model:
\begin{equation}
\label{eq:objective}
 J = - \tsav + \lambda_F \cdot F \, \, .
\end{equation}
For the gradient descent, the fit parameters $f_i$ and $a_\ell^m$ are adapted simultaneously by calculating the derivative of the objective function $J$ with respect to the parameters.
As an optimizer for the minimization problem, RMSProp \cite{Hilton2012} is used, which supports an efficient optimization for adaptive parameters of different magnitudes. 
This is realized by allowing each adaptive parameter to have a separate step size which is increased (decreased) depending on a consistent (inconsistent) direction of the gradient with respect to the parameter in two consecutive update steps.
For stability reasons, the optimizer was complemented by additional conditions for the learning rate adaption.

\subsection{Normalization}
\label{sec:normalization}

The normalization factor $C$ in equation~\eqref{eq:ellipse_equation} is generally determined by equation~\eqref{eq:normalization_likelihood} and requires a numerical integration for a non-uniform geometrical exposure $\mathcal{E} (\uvec{r})$.
Since the exposure $\mathcal{E} (\uvec{r}) = \mathcal{E} (\delta(\uvec{r}))$ depends only on the equatorial declination $\delta$ \cite{Sommers2000}, the surface integral of the term $\mathcal{E} (\delta) \, \mathcal{S}_i (\uvec{r})$ depends only on the center direction $\uvec{\Theta}_i$ of the ellipse and its relative orientation $| \alpha_i |$ towards the local spherical unit vector $\uvec{e}_\delta$ in equatorial coordinates.
As $\uvec{u}_i$ determines the orientation of the ellipse $\mathcal{S}_i (\uvec{r})$, the angle $\alpha_i$ is defined by $\cos \alpha_i = \uvec{u}_i \cdot \uvec{e}_\delta$.

\begin{figure}
\centering
\resizebox{0.49\textwidth}{!}{\includegraphics{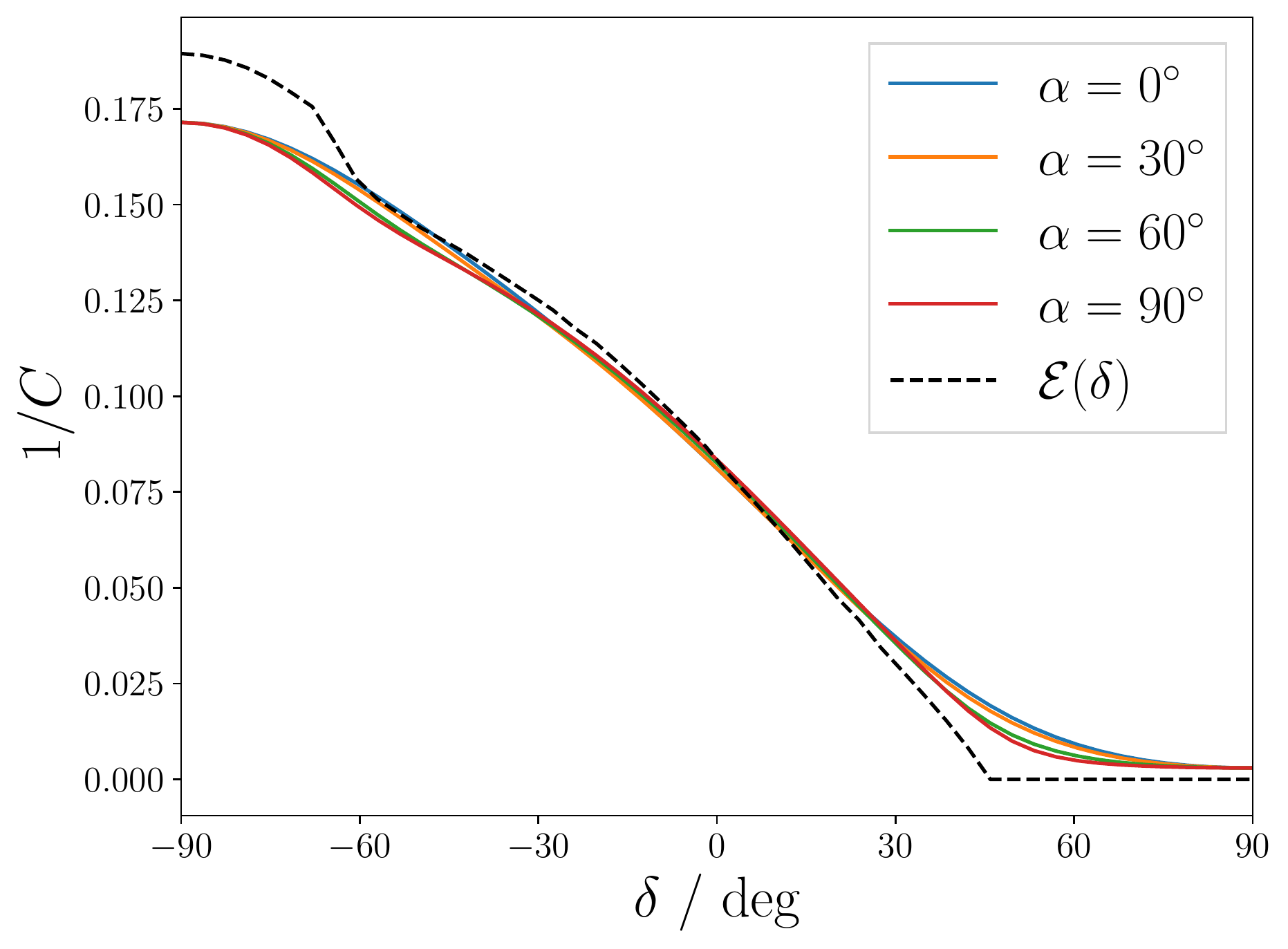}}
\caption{Integral $1/C$ of exposure-folded Gaussian probability distribution as a function of the equatorial declination $\delta$ for the center direction of the ellipse. The semi-major and semi-minor axis are $\delta_\textnormal{max}=30^\circ$ and $\delta_\textnormal{min}=10^\circ$, respectively. Colored lines denote different ellipse orientations $\alpha$. The black dashed line indicates the probability density function value of the normalized exposure $\mathcal{E}(\delta)$.}
\label{img:1_ellipse_exposure_norm}
\end{figure}

The exposure-weighted integral for an ellipse with dimensions $(\delta_\textnormal{max}, \delta_\textnormal{min}) = (30^\circ, 10^\circ)$ is visualized as a function of the equatorial declination $\delta$ of the center direction and for different values of $\alpha_i$ in Fig.~\ref{img:1_ellipse_exposure_norm}.
It can be seen that the inverse normalization factor $1/C$ approaches the probability density function value of the exposure $\mathcal{E}(\delta)$ for intermediate equatorial declinations $-60^\circ < \delta < 20^\circ$.
In the border regions of the geometrical exposure, the integral is larger than the respective exposure values due to the extent of the elliptical distribution.
For the gradient descent, the normalization factor was calculated for a grid of equatorial declinations $\delta$ and orientations $\alpha$ and then interpolated linearly between the grid points.

\section{Benchmark simulations}
\label{sec:compass_benchmark}

The sensitivity of the COMPASS fit method is evaluated in two different benchmark simulations. 
Proof of concept is given on the basis of a simple four-source model where each source contributes an equal number $N_s$ of cosmic rays.
The second benchmark extends an astrophysical simulation of an extragalactic source population \cite{Bister2020} by deflections in the GMF and the observational exposure $\mathcal{E}(\uvec{r})$. 
Therefore, given a certain source density, it provides the most reasonable estimate of the sensitivity. 
Both simulations mimic the current data set of the Pierre Auger Observatory for anisotropy studies 
(e.g. \cite{Wirtz2019b}) above energies of $40$~EeV with a total event number of $N_\textnormal{tot} = 1119$ and zenith angles up to $80^{\circ}$. \newline

\noindent\textbf{Benchmark 1: Distinct source scenario}~\newline

The first benchmark simulation consists of UHECRs with energies $E$ that follow the parameterized power law from \cite{Fenu2017} above an energy threshold of $40$~EeV.
The nuclear charges are assumed to be energy-independent and uniformly distributed between $Z=1$ and $Z=8$. 
While $N_s$ out of the $N_\textnormal{tot} = 1119$ UHECR events are assigned to each of four different, randomly placed point-like sources in the sky, the remaining cosmic rays are distributed isotropically, following the geometrical exposure $\mathcal{E}(\uvec{r})$ of the experiment.

\begin{figure}[ht]
\centering
\resizebox{0.45\textwidth}{!}{\includegraphics{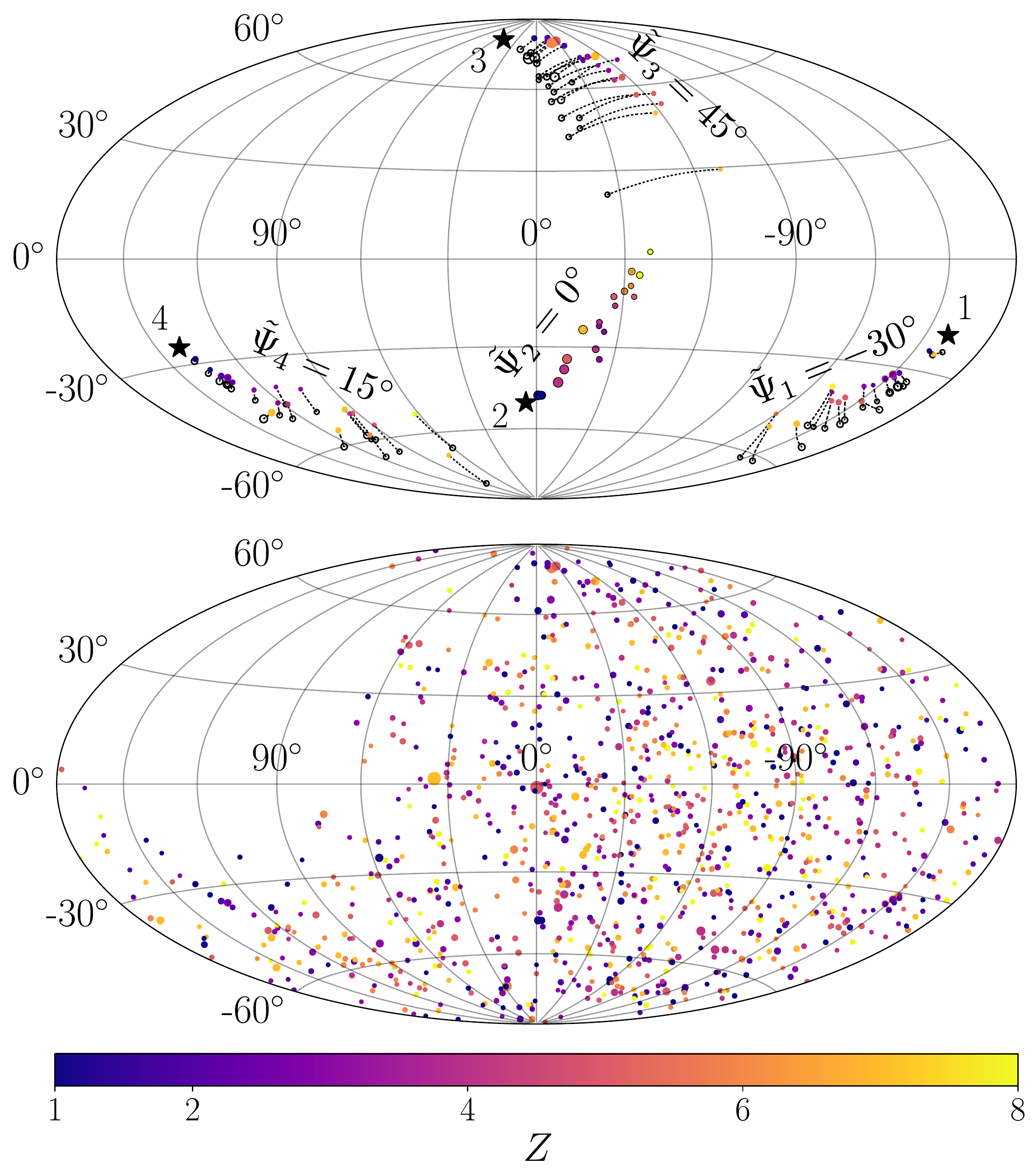}}
\caption{Construction of benchmark 1 scenario consisting of four distinct sources, in the Galactic coordinate system. Each of the four sources (black star symbols) contributes $N_s = 20$ source events with color coded charge number $Z$. (\textit{Top}) Construction of implementing GMF uncertainties for the source events. (\textit{Bottom}) Complete simulations of $80$ source events and $1039$ isotropically distributed arrival directions.}
\label{img:2_benchmark1}
\end{figure}

The deflections in the GMF are simulated as follows: 
First, the cosmic rays with magnetic rigidity $R = E / Z$ are propagated through the large-scale component of the JF12 model using a magnetic field lens \cite{Harari2000,Bretz2014}. 
Next, a rigidity-dependent Gaussian smearing of $\delta = 0.5 \times Z/E \,$[EeV]~rad is applied which corresponds approximately to the median scattering angle in the JF12 random and striated fields.
A visualization of the arrival directions of this step is shown for $N_s=20$ source events as black circles in the upper panel of Fig.~\ref{img:2_benchmark1}.

If the \textit{JF12 GMF} initialization is chosen in the fit method, we additionally apply a shift of arrival directions to mimic the uncertainties in the GMF model. 
Here, the entire pattern of source $m$ is modified by a spherical angle $\tilde{\Psi}_m$ performed by a rotation of the individual cosmic-ray arrival directions $\uvec{\Theta}_i$ from source $m$ around its direction $\uvec{r}_m$ by the angle $\tilde{\Psi}_m$.
The construction of this rotation is sketched by the dotted line in the top panel of Fig.~\ref{img:2_benchmark1}.
For the four sources, spherical angles $\tilde{\Psi}_m$ are selected in good accordance with uncertainties between existing GMF models \cite{Erdmann2016}, as examples $\tilde{\Psi}_1 = -30^{\circ}$, $\tilde{\Psi}_2 = 0^{\circ}$, $\tilde{\Psi}_3 = +45^{\circ}$ and $\tilde{\Psi}_4 = +15^{\circ}$ in order 
of the ascending Galactic longitude $l$ (from right to left in the upper panel of Fig.~\ref{img:2_benchmark1}). 
The colored symbols indicate the resulting arrival directions after this displacement.
All arrival directions, consisting of $80$ source events and $1039$ isotropic events, are shown in the lower panel of Fig.~\ref{img:2_benchmark1}. \newline

\noindent\textbf{Benchmark 2: Astrophysical simulation}~\newline

The second benchmark simulation is based on results obtained in a combined fit of the UHECR observables at the Pierre Auger Observatory~\cite{Aab2016} and their anisotropy implications for given source densities following~\cite{Bister2020}.
Here, source candidates are uniformly distributed in the universe following a source density $\rho_S$ which results in aniso-tropies due to attenuation effects during the propagation.

\begin{figure}[ht]
\centering
\resizebox{0.45\textwidth}{!}{\includegraphics{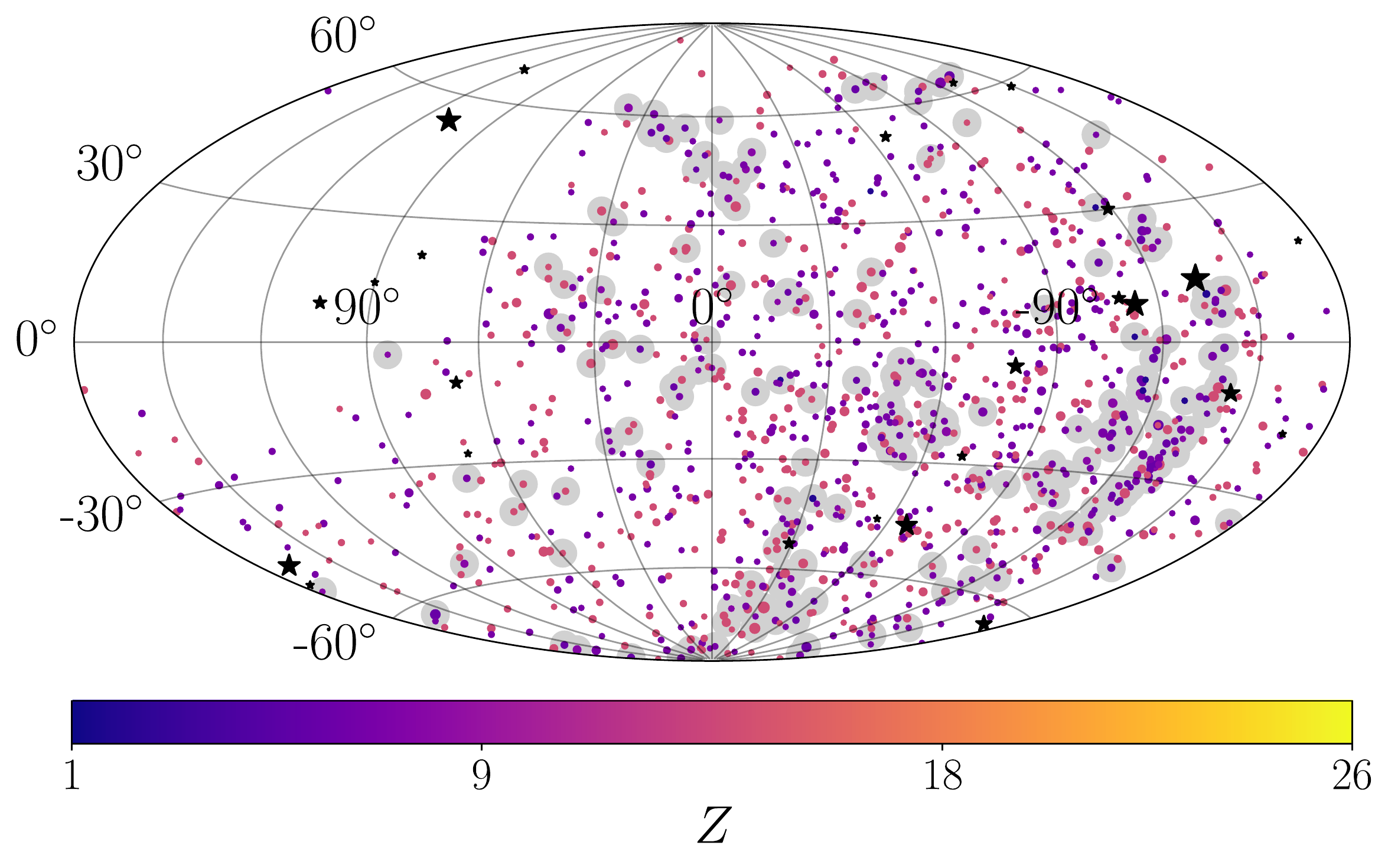}}
\caption{Example of benchmark 2 scenario consisting of an astrophysical simulation including deflection in the JF12 model for the GMF. Cosmic rays originate from uniformly distributed sources of density $\rho_S = 10^{-2}$~Mpc$^{-3}$ and are attenuated by extragalactic photon fields. Gray shaded events denote cosmic rays which originate from a source with at least three contributing events. For this specific source distribution the strongest source contributes $30$ cosmic rays, which corresponds to the median value of $1000$ simulated universes for this source density. The star symbols denote source directions where the size is proportional to the cosmic-ray contribution. The skymap is shown in the Galactic coordinate system.}
\label{img:2_benchmark2}
\end{figure}

The deflection in the GMF is applied in the same way as for the benchmark 1 scenario by using the JF12 model and a rigidity-dependent Gaussian smearing of $\delta = 0.5 \times Z/E \,$[EeV]~rad.
The relative arrival probability for different extragalactic directions and rigidities caused by the GMF (e.g.~\cite{Farrar2015}) are accounted for.
The relative observation probability resulting from the geometrical exposure of the observatory is likewise accounted for. 
Fig.~\ref{img:2_benchmark2} shows the resulting arrival directions of the benchmark 2 simulation for a source density of $\rho_S = 10^{-2}$~Mpc$^{-3}$.
The circular symbols denote cosmic-ray arrival directions, the color scale corresponds to the nuclear charge $Z$, and the gray shaded events originate from sources which contribute at least three cosmic rays. 
One can see that patterns may occur in multiple regions of the sky with strongly varying event contributions attributable mainly to the source distance.
Some sources situated outside the visible sky of the observatory (e.g. source at Galactic coordinates $l \approx 110^\circ$ and $b \approx 50^\circ$) still contribute a substantial fraction of cosmic rays due to coherent deflection in the GMF.

Additionally, if the tangent vector field is initialized as \textit{JF12 GMF}, again, an uncertainty angle $\tilde{\Psi}$ for the GMF is simulated.
To conserve consistent deflection patterns of sources in similar directions of the sky, the uncertainty $\tilde{\Psi}(\uvec{r}) = \tilde{\Psi}_a \, \uvec{r} \cdot \uvec{d}_i$ is modeled as a dipolar function with amplitude $\tilde{\Psi}_a = 45^{\circ}$ and random direction of the dipole maximum $\uvec{d}_i$ for each simulated universe $i$. 

This simulation of the UHECR universe exhibits only one single free parameter, the source density $\rho_S$, which directly determines the degree of anisotropy in the arrival directions. 
The higher the source density, the more sources are within a horizon where attenuations do not play an important role and, therefore, the more isotropic the sky is.

\section{Reconstruction of the Galactic magnetic field}
\label{sec:compass_reconstruction}

In this section, proof of concept is provided by showing that the orientation of patterns can be correctly reconstructed based on the benchmark 1 simulation from section~\ref{sec:compass_benchmark}.

During the minimization process, the modification angle $\Psi(\vartheta, \varphi)$ rotates the ellipses of the signal hypothesis of equation~\eqref{eq:psi} such that they align with elongated patterns in the cosmic-ray arrival direction distribution. 
For the \textit{JF12 GMF} initialized vector field $\uvec{u}_0$, the angle $\Psi(\vartheta, \varphi)$ corresponds directly to a correction of the JF12 model in sky regions where a significant pattern is found.
Hence, for the benchmark 1 simulation, the final angle $\Psi_i \equiv \Psi(\vartheta_i, \varphi_i)$ of cosmic rays $i$ which originate from one of the simulated sources $m$ is expected to approach the simulated uncertainty $\tilde{\Psi}_m$.
Here, the necessary correction is modeled by the spherical angles $\tilde{\Psi}_m$ for the four sources where $\tilde{\Psi}_1 = -30^{\circ}$, $\tilde{\Psi}_2 = 0^{\circ}$, $\tilde{\Psi}_3 = +45^{\circ}$ and $\tilde{\Psi}_4 = +15^{\circ}$ are chosen (cf. section~\ref{sec:compass_benchmark}).
Note that a non-linear deflection behavior in the GMF may disturb the correct values of $\tilde{\Psi}_m$.
This effect is particularly strong for cosmic rays with a low rigidity $E_i/Z_i$, i.e. for high absolute deflection angles with respect to their source.

Here, for the first application of the fit, the order of the spherical harmonics expansion of equation~\eqref{eq:psi} is defined as $k = 5$, which corresponds to $36$ free fit parameters.
In this case, modifications of the GMF model can be performed coherently in sky regions that have angular scales above the order of $180^\circ / k = 36^\circ$.
The degree of modification $\Psi$ itself is constrained by the hyperparameter in the objective function~\eqref{eq:objective} where a value of $\lambda_F = 1$ is chosen for this purpose.
For the ellipse geometry, values of ($\delta_\textnormal{max}, \, \delta_\textnormal{min}) = (10^\circ, \, 5^\circ$) are chosen in equation~\eqref{eq:ellipse_equation}.
Furthermore, the Gaussian reference model from equation~\eqref{eq:reference_likelihood_gaussian} was selected for the likelihood ratio where the effective Gaussian width is $\sqrt{10^\circ \times 5^\circ} \approx 7.1^\circ$.

\begin{figure}
\centering
\resizebox{0.45\textwidth}{!}{\includegraphics{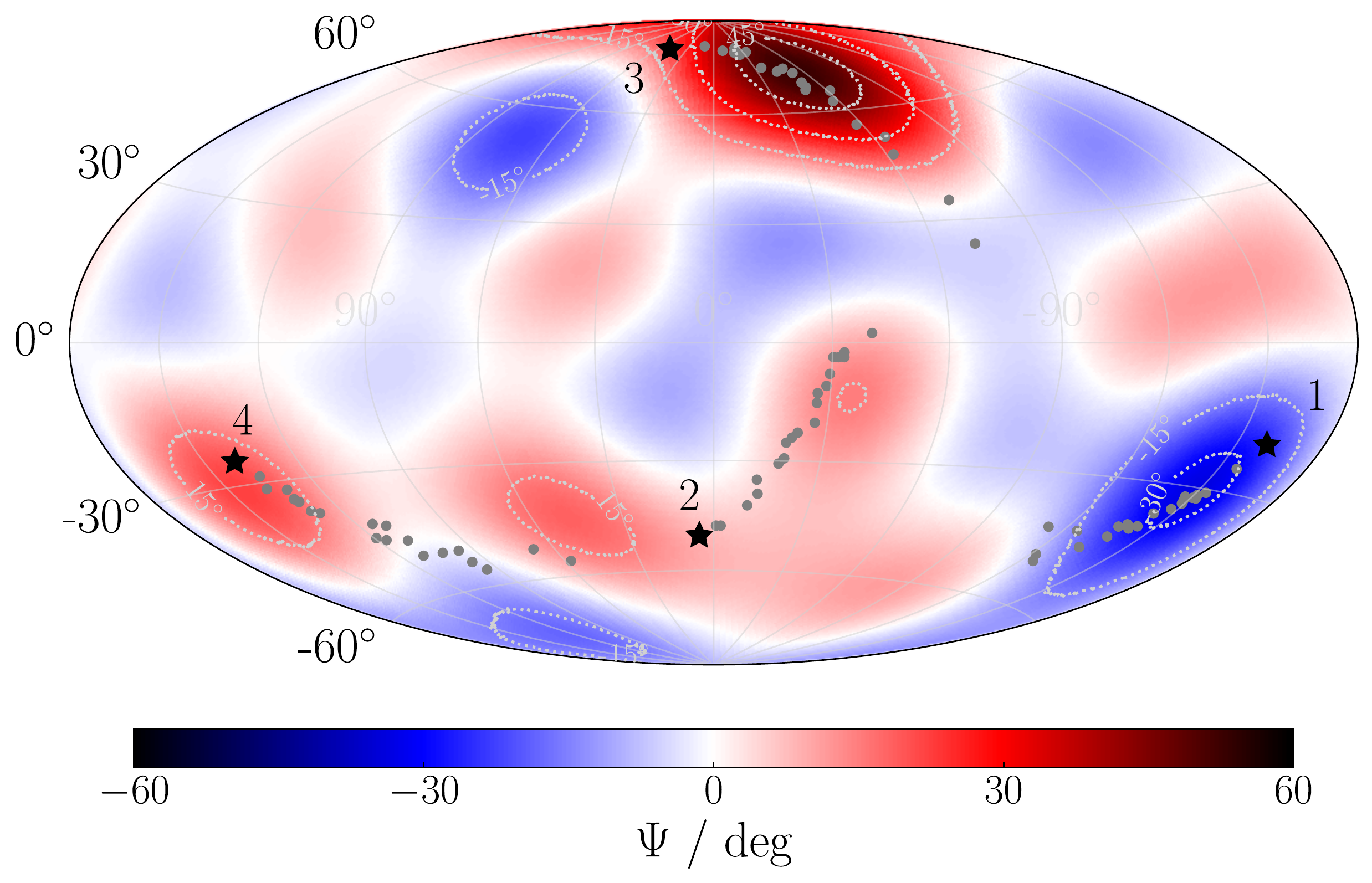}}
\caption{Visualization of the modification angles $\Psi(\vartheta, \varphi)$ (color-coded) which is defined relative to the JF12 model (\textit{JF12 GMF} initialization for $\uvec{u}_0$) after a fit to benchmark 1 scenario with $N_s = 20$ source events. Gray circles denote cosmic rays that originate from one of the four sources and the dotted lines mark contours of $15^{\circ}$-spacing in $\Psi$.}
\label{img:3_gmf_reconstruction}
\end{figure}

The fitted modification function $\Psi(\theta, \varphi)$ is visualized in Fig.~\ref{img:3_gmf_reconstruction} together with the cosmic rays that originate from the simulated source candidates.
As an overall impression the color code in the vicinity of the source candidates $m$ agrees with the simulated  uncertainties $\tilde{\Psi}_m$. 
To quantify the method's reconstruction abilities, for each source $m$ the fitted $\Psi_i$ for the $10$ closest cosmic rays that originate from the source are averaged.
The corresponding averaged values are $\langle \Psi_1 \rangle = (-32 \pm 1)^{\circ}$, $\langle \Psi_2 \rangle = (6 \pm 2)^{\circ}$, $\langle \Psi_3 \rangle = (+47 \pm 5)^{\circ}$ and $\langle \Psi_4 \rangle = (+15 \pm 5)^{\circ}$, which are in good agreement with the simulated uncertainties $\tilde{\Psi}_m$.

The next step is to investigate if orientations of patterns as simulated with the JF12 model can also be captured without including information on the explicit GMF model.
For this purpose, we chose the \textit{Galactic meridians} initialization of $\uvec{u}_0$ where initial ellipse orientations are aligned with the local spherical unit vector $\uvec{e}_\vartheta$ of the latitude in the Galactic coordinate system.
Since no information on the simulated GMF is included, the penalization factor of equation~\eqref{eq:objective} is not required and is therefore set to $\lambda_F = 0$.
Thus, the tangent vector field can be rotated by the angle $\Psi(\theta, \varphi)$ without constraint.
For this setup, the degree of the spherical harmonics expansion was decreased to $k = 4$ to avoid rapid changes of $\Psi$ on small angular scales.
The order $k$ of the spherical harmonics expansion is the most challenging free parameter since the optimal choice depends on the angular scale of domains with a coherent GMF deflection.
While more complex patterns can generally be fitted with an increasing order of $k$, these structures are more difficult to interpret.

\begin{figure}
\centering
\resizebox{0.45\textwidth}{!}{\includegraphics{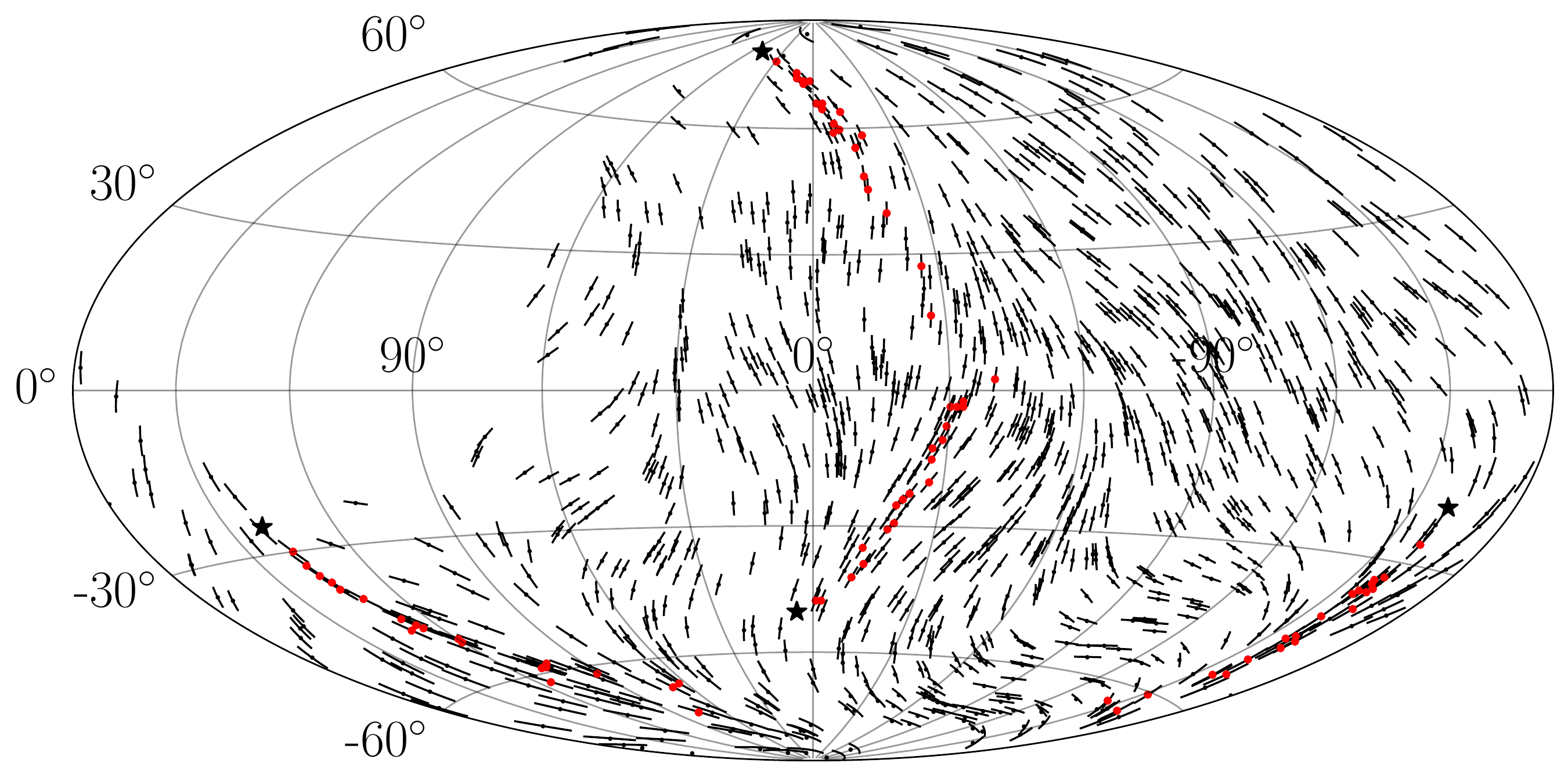}}
\caption{Visualization of the tangent vector field $\uvec{u}(\Theta_i)$ (black lines) of equation~\eqref{eq:vector_field} using the \textit{Galactic meridians} initialization of $\uvec{u}_0 (\uvec{r})$. Black star symbols show the simulated source directions while the red circular symbols mark the $N_s = 20$ events per source which have been displaced by the JF12 model without simulated GMF uncertainties.}
\label{img:3_pattern_reconstruction}
\end{figure}

A visualization of the fitted tangent vector field $\uvec{u}(\Theta_i)$ for each individual cosmic-ray arrival direction $\Theta_i$ is presented in Fig.~\ref{img:3_pattern_reconstruction} where the source events are highlighted in red.
All four deflection patterns in this simulation were successfully captured by an alignment of the tangent vector field  $\uvec{u}(\uvec{r})$ along the local track of source events.
There is also structure visible in sky regions without source contribution where fluctuations of isotropically distributed arrival directions were connected along their most prominent patterns.
In the sky region at Galactic coordinates $(l, b) \approx (-90^\circ, -75^\circ)$ the isotropic fluctuation was even strong enough to rotate the initialized tangent vector field $\uvec{u}_0$ by up to $95^\circ$.
This suggests that arbitrarily oriented alignment patterns of cosmic-ray arrival directions can be captured even when oriented orthogonally with respect to the chosen initialization $\uvec{u}_0$.

To assess whether a fitted pattern is caused by isotropic fluctuations, the individual cosmic-ray test statistics from equation~\eqref{eq:individual_test_statistic} have to be compared to those found in isotropic skies.
These sensitivity studies are presented in section~\ref{sec:compass_sensitivity}.

\section{Reference model of the likelihood ratio}
\label{sec:compass_reference_model}

In this section, two different choices of the reference model (cf. section \ref{sec:strategy}) for the likelihood ratio as defined in equation~\eqref{eq:individual_test_statistic} are studied: 
the isotropic model $\mathcal{E}$ which follows the geometrical exposure of an experiment, and the Gaussian model $\mathcal{G}_i$ with identical signal contributions $f_i$ as assigned to the elliptically shaped signal model $\mathcal{S}_i$.
Again, the ellipse geometry is defined as ($\delta_\textnormal{max}, \, \delta_\textnormal{min}) = (10^\circ, \, 5^\circ$), the initialization for $\uvec{u}_0$ is \textit{Galactic meridians}, the spherical harmonics order is $k = 4$, and the hyperparameter $\lambda_F = 0$.
To obtain an impression of the performance over the sky, it is useful to investigate the individual test statistics $\tsi$ defined in equation~\eqref{eq:individual_test_statistic} as well as the anticipated signal contribution $f_i$ from equation~\eqref{eq:signal_likelihood}. \newline

\noindent\textbf{Isotropic reference model $\mathcal{E}$}~\newline

\begin{figure*}[ht]
\centering
\resizebox{\textwidth}{!}{\includegraphics{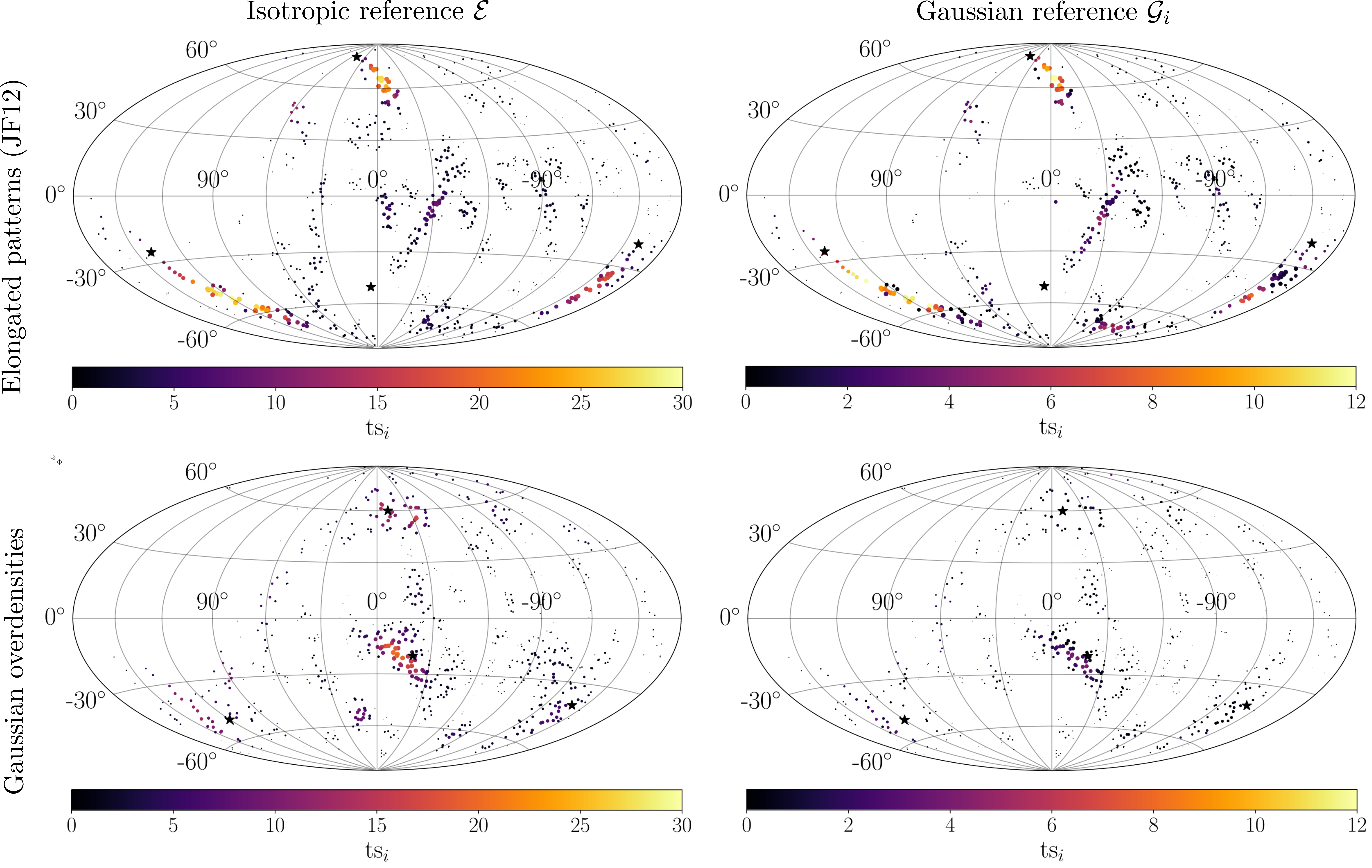}}
\caption{Comparison of fit performance for (\textit{left}) isotropic $\mathcal{E}$ and (\textit{right}) Gaussian reference model $\mathcal{G}_i$, in Galactic coordinates. (\textit{Top}) Benchmark 1 simulation from section~\ref{sec:compass_benchmark} with $N_s=20$ events per source and deflection in the JF12 model. (\textit{Bottom}) Gaussian overdensity with a cluster of $N_s = 20$ events distributed around the star symbols with a Gaussian width of $10^\circ$ and without GMF deflection. The color code of each event $i$ corresponds to the individual test statistic $\tsi$. The size of the circular symbols is proportional to the signal contribution $f_i$ with identical normalization among all four figures.}
\label{img:4_gaussian_exposure_comparison}
\end{figure*}

The resulting test statistics $\tsi$ after the fit using the isotropic reference model (cf. equation~\eqref{eq:reference_likelihood_isotropy}) is shown in the left panel of Fig.~\ref{img:4_gaussian_exposure_comparison}. 
The fit results of the benchmark 1 simulation are displayed in the upper panel with $N_s = 20$ source events in each of the four patterns.
The fitted signal contribution $f_i$ for the events $i$ are proportional to the size of the circular symbols where a common normalization among all four figures is chosen.

Clearly, for benchmark 1 scenario in the upper panel, the patterns produced by the four simulated sources exhibit cosmic rays with a substantially larger test statistic $\tsi$ compared to the isotropically distributed background events. 
Accordingly, the anticipated signal contribution $f_i$ of the source events is larger, as shown by the size of the markers.
Some local clusters of events with test statistics $\tsi > 0$ can also be found in the isotropically distributed background events, however, with considerably smaller values of both the test statistic $\tsi$ and the signal contribution $f_i$.
The average test statistic from equation~\eqref{eq:test_statistic} is $\tsav \approx 1.6$. 
The highest signal fractions reach values of about $f_i=1.4 \%$, which is in good agreement with the $20$ of $1119$ injected signal cosmic rays per source. 
Note that the complete signal contribution of $20 / 1119 \approx 1.8 \%$ is not necessarily reached even for the innermost cosmic ray of the pattern due to fluctuations in the isotropic background and the ellipse's limited extent of $10^{\circ}$ in the semi-major axis, which is mostly less than the extent of the pattern.

To assess the impact of solely overdense but not elongated structures on the test statistic, a new simulation is studied which again consists of four sources each emitting $20$ cosmic rays. 
Instead of simulating deflections in the GMF, the source events are drawn from a Fisher distribution \cite{Fisher1953} of a width of $10^{\circ}$ centered on the direction of the source.
Here, to enable a better comparison between both scenarios, the source directions were approximately centered within the resulting arrival patterns of the benchmark 1 simulation.

As shown in the lower panel of Fig.~\ref{img:4_gaussian_exposure_comparison}, the method also responds with high individual test statistics $\tsi$ and anticipated signal contributions $f_i$ due to the event excess of $N_s = 20$ relative to an isotropic expectation. 
However, for three of the four Fisher distributions the resulting test statistics are much smaller than in the case of the benchmark 1 simulation. \newline

\noindent\textbf{Gaussian reference model $\mathcal{G}_i$}~\newline

The right panel of Fig.~\ref{img:4_gaussian_exposure_comparison} again shows the individual test statistic $\tsi$ (color coded) and the fitted signal contributions $f_i$ (size of circles) for the Gaussian reference hypothesis $\mathcal{G}_i$ as defined in equation~\eqref{eq:reference_likelihood_gaussian}.
For the benchmark 1 simulation in the upper panel, the anticipated signal contribution $f_i$ is approximately equal to the case where the isotropic reference model $\mathcal{E}$ was chosen, as can be estimated from the size of the markers.
However, while in the case of the isotropic reference model both the event excess and the elongation of the structure contributed to the test statistic, for the Gaussian reference model only the latter information can be used.
Thus, on the one hand, the overall scale of the individual test statistics $\tsi$ is much smaller, as reflected by the color scale.
Therefore, the average test statistic of equation~\eqref{eq:test_statistic} drops to $\tsav \approx 0.8$.
On the other hand, since there is no sensitivity to overdense regions, some of the patterns that were caused by fluctuations of isotropically distributed background events are no longer visible.
Thus, the purity of detected patterns is increased compared to when the isotropic reference model was used.

Again, the response to Gaussian overdensities is assessed in the bottom panel of Fig.~\ref{img:4_gaussian_exposure_comparison} with four Fisher-distributed event clusters of $10^\circ$ Gaussian width. 
Since the Gaussian-shaped event structures are well described by the Gaussian reference hypothesis $\mathcal{G}_i$, there is a significant loss in the test statistic for the overdense sky regions compared to when the isotropic reference $\mathcal{E}$ is used.
In the vicinity of three Gaussian event clusters, there is only barely more fitted signal contribution $f_i$ compared to the remaining sky.
The individual test statistics $\tsi$ visibly deviates from natural isotropic fluctuations only for the events of one of the Gaussian clusters, namely at coordinates $(l, b) = (-20^\circ, -20^\circ)$. 

As there are already known event excesses in UHECR data, e.g. in data of the Pierre Auger Observatory for this energy threshold (e.g. \cite{Aab2018a,Caccianiga2019}), there is a risk of detecting these features again rather than new elongated structures when using the isotropic reference model $\mathcal{E}$.
Therefore, in the following we use the Gaussian-like reference model $\mathcal{G}_i$ where the effects of overdensities are mostly canceled out by the likelihood ratio.

\section{Sensitivity studies}
\label{sec:compass_sensitivity}

In this section we investigate the sensitivity of the COMPASS method with respect to its ability to reject isotropic distributions of cosmic-ray arrival directions.
According to the findings from section~\ref{sec:compass_reference_model}, for the following subsections the Gaussian reference model $\mathcal{G}_i$ (cf. equation~\eqref{eq:reference_likelihood_gaussian}) is chosen in the likelihood ratio. 
In addition, following section~\ref{sec:compass_reconstruction} and the studies in section~\ref{sec:compass_sensitivity_hyperpar}, the tangent vector field $\uvec{u}_0$ is initialized along the \textit{Galactic meridians} --- i.e. $\uvec{u}_0$ is equal to the local spherical unit vector $\uvec{e}_\theta$.
Therefore, the penalization term $F$ in equation~\eqref{eq:objective} is removed by setting $\lambda_F = 0$.
For a comparison of the sensitivity with a more classical analysis to search for elongated structures refer to \cite{Wirtz2020}.

\begin{figure}
\centering
\resizebox{0.45\textwidth}{!}{\includegraphics{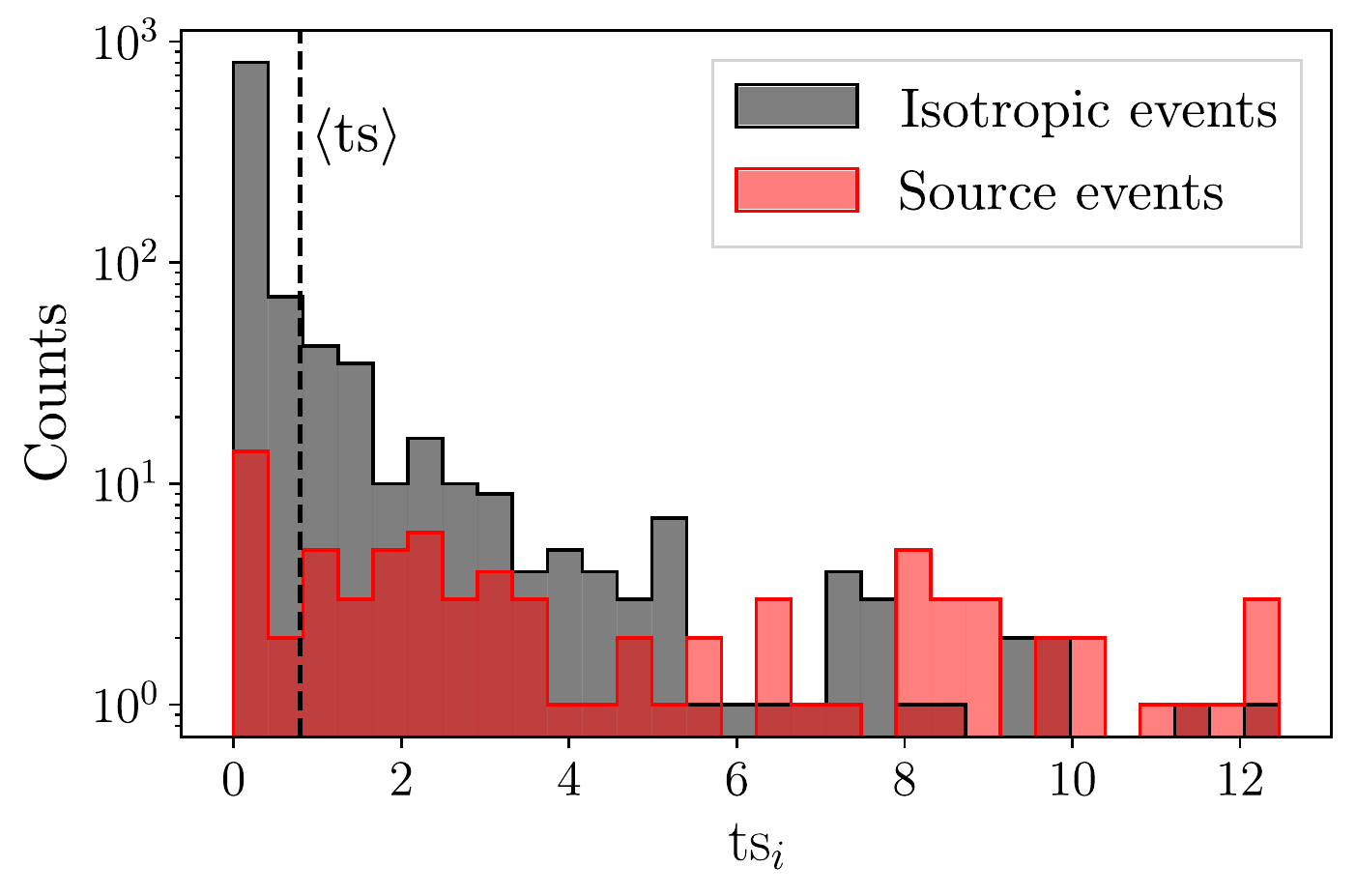}}
\caption{Distribution of individual test statistics $\tsi$ as obtained in the benchmark 1 simulation (cf. upper right panel of Fig.~\ref{img:4_gaussian_exposure_comparison}). The red (gray) histogram shows the contribution of the $80$ source (isotropically distributed) events. The dashed vertical black line denotes the average test statistic $\tsav$ of all events in the sky.}
\label{img:5_individual_test_statistic}
\end{figure}

\subsection{Sensitivity for distinct source scenario}

The distribution of individual test statistics $\tsi$ as obtained in the benchmark 1 simulation from section~\ref{sec:compass_benchmark} is presented in Fig.~\ref{img:5_individual_test_statistic}.
As already suggested by the upper right panel of Fig.~\ref{img:4_gaussian_exposure_comparison}, most of the events that exhibit high test statistics are attributed to one of the four sources.
In total, more than half of the source events show test statistics larger than $3$, which, in turn, is only reached for about $5 \%$ of the isotropic events.
Instead, the isotropic distribution peaks close to zero, with about $50 \%$ of events exhibiting test statistics smaller than $10^{-5}$.
One can of course find a statistical measure to reject isotropy based on the evaluation of events with a high test statistic, i.e. based on the tail of the distribution in Fig.~\ref{img:5_individual_test_statistic}.
However, it was found that the average test statistic $\tsav$ provides the most stable measure for various simulation setups.

\begin{figure}
\centering
\resizebox{0.45\textwidth}{!}{\includegraphics{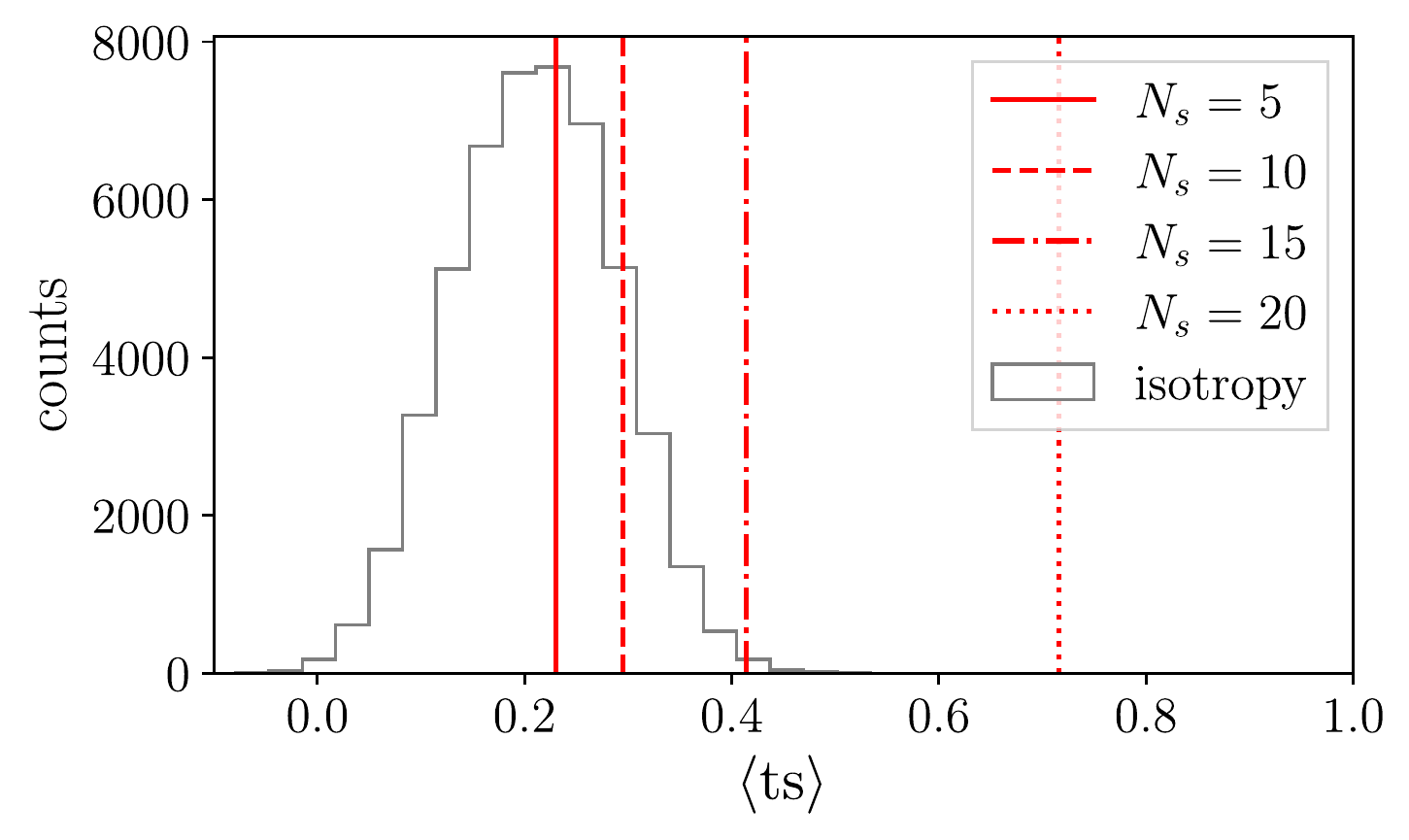}}
\caption{Distribution of average test statistic $\tsav_\textnormal{iso}$ for $10^4$ isotropic skies (gray) compared to the values obtained in the benchmark 1 simulation (vertical red lines) with $N_s = 5$ (solid), $N_s = 10$ (dashed), $N_s = 15$ (dash-dotted), and $N_s = 20$ (dotted) source events.}
\label{img:5_sensitivity_benchmark1}
\end{figure}

In the next step, we evaluated the average test statistic $\tsav$ for different numbers of source events $N_s = 5, \, 10, \, 15,$ $\, 20$ in the benchmark 1 simulation.
As shown in Fig.~\ref{img:5_sensitivity_benchmark1}, the resulting values for the average test statistics are $\tsav = 0.23, \, 0.30, \, 0.41, \, 0.72$, respectively. 
To calculate the chance probability $p_\textnormal{val}$ of obtaining these average test statistics from an isotropic arrival-direction distribution, the method is additionally applied to $5 \times 10^4$ isotropic realizations of the sky which follow the geometrical exposure of the observatory. 
The distribution of the average test statistics $\tsav_\textnormal{iso}$ for isotropic skies is shown as a gray histogram in Fig.~\ref{img:5_sensitivity_benchmark1}.
While there is no isotropic sky yielding a higher average test statistic than the scenarios with $N_s = 20$ source events, the isotropic chance probabilities for the simulations with smaller source events are $p_\textnormal{val} = 0.40, \, 0.14, \, 0.003$, in order of increasing $N_s$. 

As expected from the central limit theorem (cf. section~\ref{sec:strategy}), the gray histogram shows that the average test statistic $\tsav_\textnormal{iso}$ for an isotropic arrival sky approximately follows a Gaussian distribution.
Thus, the sensitivity for the scenario shown in Fig.~\ref{img:2_benchmark1} with $N_s = 20$ source events can be estimated by fitting a Gaussian distribution to the gray histogram.
In this case, the estimated chance probability is about $3 \times 10^{-11}$ which translates to about $6.5 \, \sigma$ standard deviations in the normal distribution.

\subsection{Sensitivity for the astrophysical universe}

\begin{figure}
\centering
\resizebox{0.45\textwidth}{!}{\includegraphics{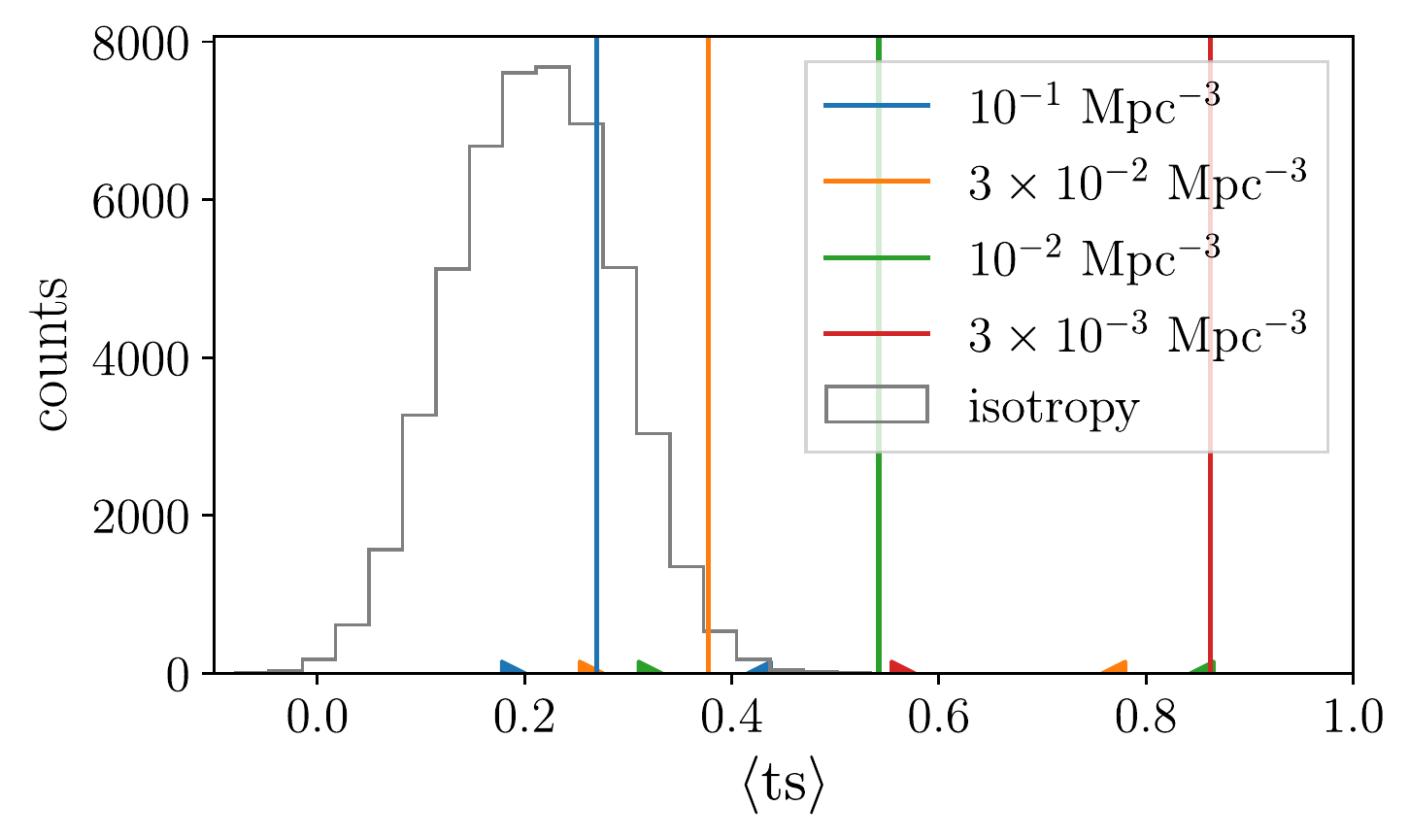}}
\caption{Distribution of average test statistics $\tsav_\textnormal{iso}$ for $5 \times 10^4$ isotropic skies (gray) compared with the values obtained in the benchmark 2 simulation (vertical lines) with source densities of $\rho_S = 10^{-1} /\text{Mpc}^3$ (blue), $\rho_S = 3 \times 10^{-2} /\text{Mpc}^3$ (orange), $\rho_S = 10^{-2} /\text{Mpc}^3$ (green), and $\rho_S = 3 \times 10^{-3} /\text{Mpc}^3$ (red). The small triangles denote the $68\%$ quantiles.}
\label{img:5_sensitivity_benchmark2}
\end{figure}

While the previous section provided an idea of the sensitivity for comparably clear patterns with a certain signal contribution, this section evaluates the expected implication for an astrophysical universe of uniformly distributed UHECR sources.
For the benchmark 2 simulations, $300$ simulated universes for each of the source densities $\rho_S = (10^{-1}, \, 3 \times 10^{-2}, \, 10^{-2}, \, 3 \times 10^{-3})$~Mpc$^{-3}$ were investigated.
The average test-statistic distribution as obtained from the fit exhibits a comparably large spread, which is consistent with the fluctuations in the degree of anisotropy.
The median and $68$ percentiles of the average test statistics for the four source densities are $\tsav = 0.27^{+0.16}_{-0.08}, \, 0.38^{+0.39}_{-0.11}, \, 0.54^{+0.31}_{-0.22},\, 
0.86^{+1.92}_{-0.30}$, respectively, as visualized in Fig.~\ref{img:5_sensitivity_benchmark2}. 
As expected, the test statistic increases with decreasing source density as the arrival scenarios become increasingly anisotropic. 

For the isotropic chance probability $p_\textnormal{val}$, the average test statistic is again compared to the fit results for the $5 \times 10^4$ isotropic realizations which are shown as a gray histogram in Fig.~\ref{img:5_sensitivity_benchmark2}.
For the source densities of $10^{-1}$~Mpc$^{-3}$ and $3 \cdot 10^{-2}$~Mpc$^{-3}$ the isotropic chance probability can be directly determined by the fraction of the gray distribution that is above the corresponding test-statistic values.
Here, the chance probabilities yield $p_\textnormal{val} = 0.229, \, 0.013$ for the two source densities respectively.
For the smaller source densities of $10^{-2}$~Mpc$^{-3}$ and $3 \times 10^{-3}$~Mpc$^{-3}$, the isotropic chance probability can again be estimated by parameterizing the null hypothesis with a Gaussian distribution.
In this case, the estimated values are $8.3 \times 10^{-6}$ and $1.4 \times 10^{-17}$, respectively, which correspond to a deviation of $4.3 \, \sigma$ and $8.5 \, \sigma$ standard deviations in the normal distribution.
Thus, in the case of a result on data that is compatible with an isotropic distribution, the density of UHECR sources for this astrophysical model can be estimated.

\begin{figure}
\centering
\resizebox{0.45\textwidth}{!}{\includegraphics{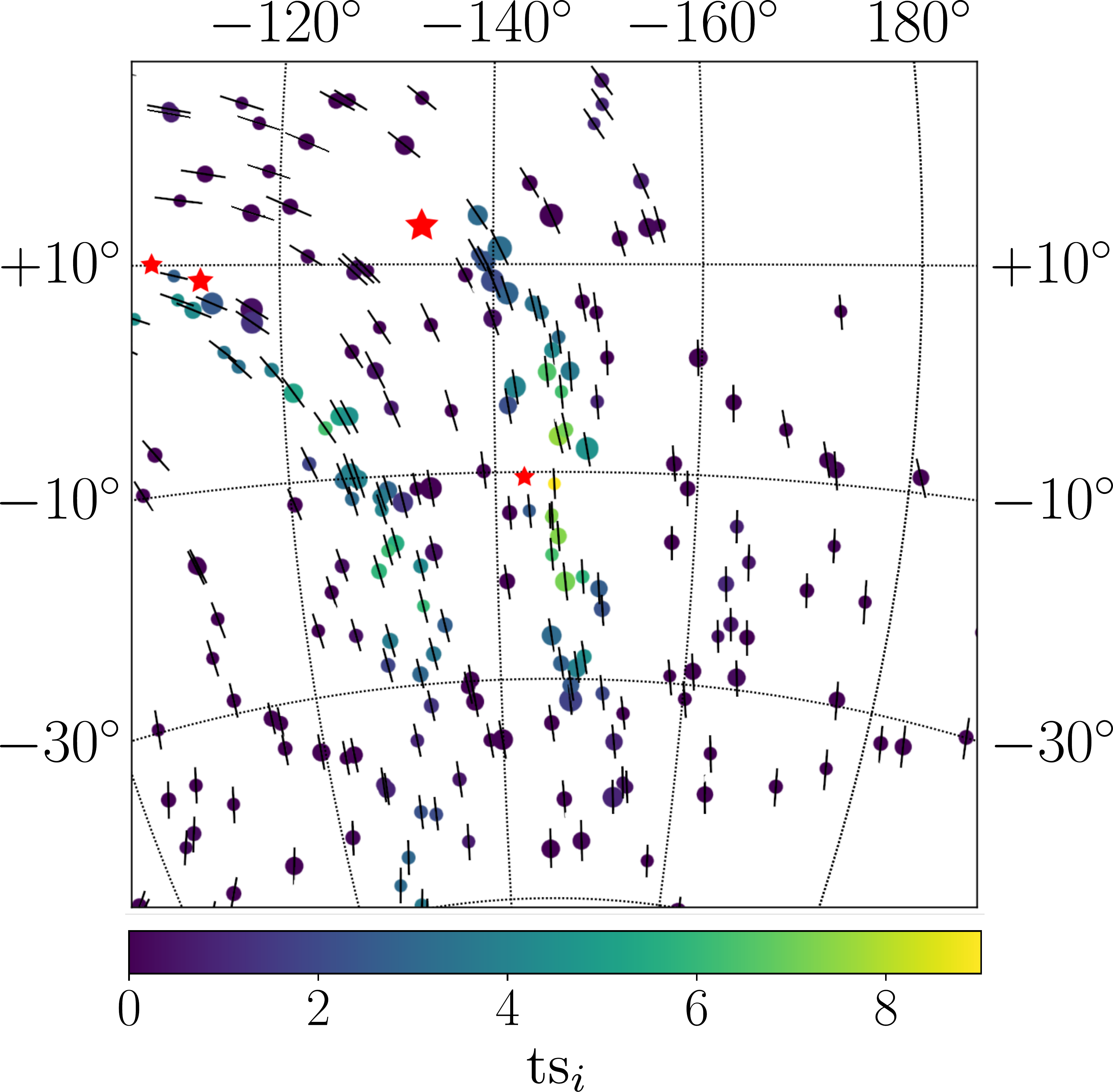}}
\caption{Example realization of arrival directions from the benchmark 2 simulation with a source density of $\rho_S = 10^{-2}$~Mpc$^{-3}$. The strongest of the sources (red star symbols) at Galactic longitude $l \approx -133^{\circ}$ and Galactic latitude $b \approx 14^{\circ}$ contributes with $29$ cosmic rays. Sizes of the cosmic-ray events (colored circles) correspond to the energy $E$, the color code to the individual test statistics $\tsi$, and the short black lines indicate the tangent vector field $\uvec{u}_0$, i.e. the orientation of the major-axis of the ellipses.}
\label{img:5_pattern_benchmark2}
\end{figure}

Fig.~\ref{img:5_pattern_benchmark2} shows arrival directions in the sky region around the strongest individual test statistic $\tsi$ for the scenario that exhibits the median average test statistic $\tsav$ out of the $300$ simulations with a source density of $10^{-2}$~Mpc$^{-3}$. 
Here, the strongest individual test statistic is $\tsi = 8.6$ and the corresponding cosmic-ray event (yellow point in the center of the sky patch) is part of the pattern from the source at Galactic coordinates $l \approx -133^\circ$ and $b \approx 14^\circ$. 
Contributing with a total of $29$ events, this is the strongest source in this realization.
The orientation of the tangent vector field $\uvec{u}(\vartheta, \varphi)$ (short black lines) additionally suggests that the alignment works reasonably well even for patterns that are only separated by about $20^\circ$.
Thus, the vector field $\uvec{u}(\vartheta, \varphi)$ is expected to provide an adequate coherent description of the deflection in the GMF for sufficiently strong signals.

\subsection{Optimization of free parameters}
\label{sec:compass_sensitivity_hyperpar}

In this subsection we evaluate the impact of the free parameters more profoundly based on the astrophysical benchmark 2 simulation with a source density of $\rho_s = 3 \times 10^{-2}$~Mpc$^{-3}$ from section~\ref{sec:compass_benchmark}.
Firstly, the initialization method of the tangent vector field $\uvec{u}_0$ and accordingly the free parameter $\lambda_F$ are addressed.
Secondly, the impact of the ellipse geometry, namely the semi-major and semi-minor axes, on the performance is studied. \newline

\subsubsection{Confidence in the \textit{JF12 GMF} initialization}

The initialization of the tangent vector field $\uvec{u}_0$ according to the predictions from the JF12 model (\textit{JF12 GMF}) is visualized in the top panel of Fig.~\ref{img:vector_field}.
As pointed out in section~\ref{sec:strategy}, depending on the reliability of the GMF model it may be beneficial to constrain the allowed deviations $\Psi (\vartheta, \varphi)$ with equation~\eqref{eq:gmf_objective}, since this reduces high test statistics from fluctuations in isotropic arrival distributions.
Therefore, the isotropic chance probability $p_\textnormal{val}$ is investigated as a function of the free objective parameter $\lambda_F$ in equation~\eqref{eq:objective} for two reasonable estimates of the uncertainties in GMF models.

\begin{figure}
\centering
\resizebox{0.45\textwidth}{!}{\includegraphics{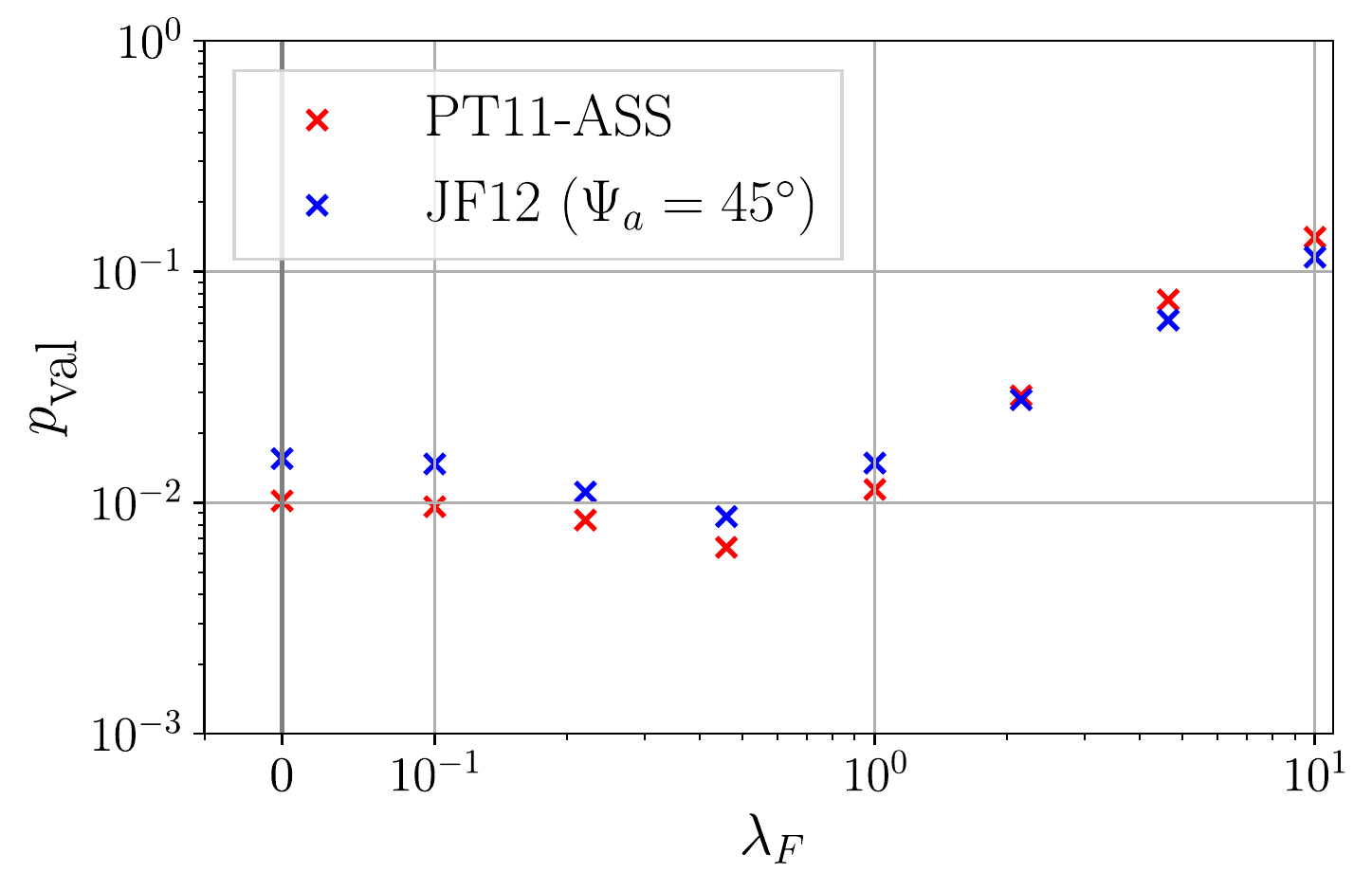}}
\caption{Scan of hyperparameter $\lambda_F$ for the confidence in the assumed GMF model  (here JF12). The initialization of the tangent vector field $\uvec{u}_0$ follows the \textit{JF12 GMF} procedure and the isotropic chance probability $p_\textnormal{val}$ is derived for the benchmark 2 simulation with a source density of $\rho_S = 3 \times 10^{-2}$~Mpc$^{-3}$ from section~\ref{sec:compass_benchmark}. The red crosses show deflections in the PT11-ASS model and the blue crosses show deflections in the JF12 model, which is modified by a dipolar modulated uncertainty angle with amplitude $\Psi_a$.}
\label{img:5_lambda}
\end{figure}

The first estimate is obtained by simulating the deflection in the GMF with the model of Pshirkov et al. using an antisymmetric disk field (PT11-ASS)~\cite{Pshirkov2011} instead of the JF12 model used in section~\ref{sec:compass_benchmark}.
The second estimate is given by a modification of the JF12 model with dipolar distributed modification angles of amplitude $\Psi_a=45^\circ$ as described in section~\ref{sec:compass_benchmark}.
The isotropic chance probabilities for both estimates are presented in Fig.~\ref{img:5_lambda} as a function of $\lambda_F$.
For high values of $\mathcal{O}(\lambda_F) = 10$, the tangent vector field is too stiff and the test statistic is therefore obtained for ellipses which are not aligned with the simulated structures.
There is a minimum for both assumptions of GMF uncertainties located consistently at values about $\lambda_F \approx 0.5$.
For an entirely flexible tangent vector field, i.e. for the parameter $\lambda_F = 0$, the additionally found patterns from isotropic skies reduce the sensitivity slightly; however, the overall isotropic chance probability is still of the same order.

Since the uncertainties of the GMF models might be even higher than assumed here and particularly uncertain in the Galactic disk region, the advantage of a hyperparameter $\lambda_F > 0$ may be even smaller.
Therefore, in the following ellipse geometry investigation the penalization term $F$ is canceled in equation~\eqref{eq:objective} and the \textit{Galactic meridians} initialization is utilized which features a symmetry with respect to the Galactic disk. \newline

\subsubsection{Ellipse geometry}

Here, we assess the impact of the ellipse geometry, the semi-major axis width $\delta_\textnormal{max}$, and the semi-minor axis width $\delta_\textnormal{min}$, for the astrophysical benchmark 2 simulation.
For the two-dimensional scan of the widths, angular bins of ($3^\circ$, $5^\circ$, $7^\circ$, $10^\circ$, $15^\circ$, $20^\circ$) were chosen where the condition $\delta_\textnormal{max} > \delta_\textnormal{min}$ is required by design.
Thus, there are $15$ different scanned ellipse geometries.
To calculate the isotropic chance probability $p_\textnormal{val}$, the analysis is applied to a total of $10^4$ isotropic skies for each geometry.

\begin{figure}
\centering
\resizebox{0.45\textwidth}{!}{\includegraphics{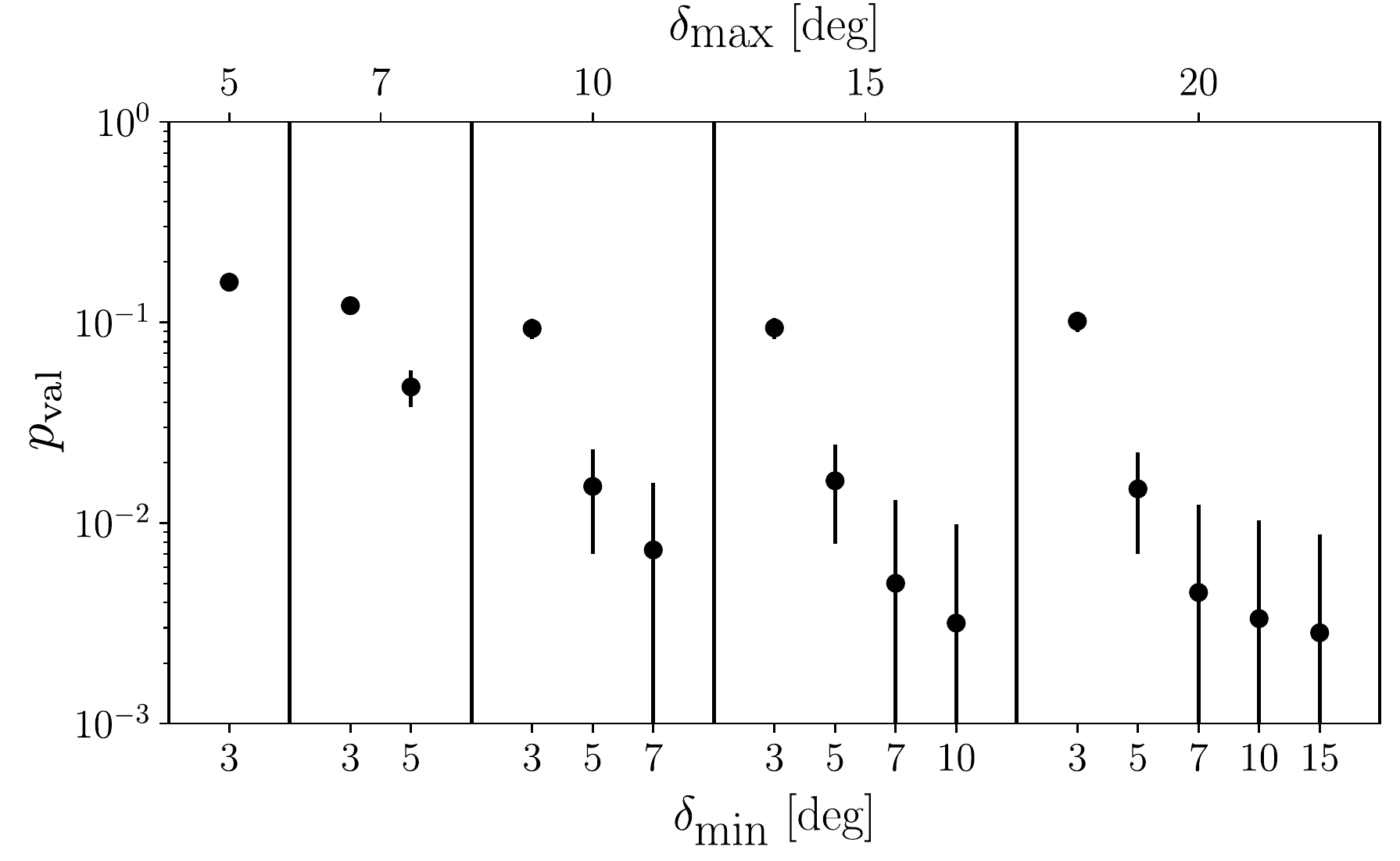}}
\caption{Scan of the ellipse geometry ($\delta_\textnormal{max}$, $\delta_\textnormal{min}$) for the \textit{Galactic meridians} initialization and $\lambda_F = 0$ evaluated on the astrophysical benchmark 2 simulation with source density of $\rho_S = 3 \times 10^{-2}$~Mpc$^{-3}$. The isotropic chance probability is calculated with a total of $10^4$ isotropic skies for each of the ellipse constellations.}
\label{img:5_rmax_rmin}
\end{figure}

The resulting median chance probabilities $p_\textnormal{val}$ of $300$ sky realizations are displayed in Fig.~\ref{img:5_rmax_rmin} where each of the five segments indicate one of the semi-major axes $\delta_\textnormal{max}$.
Generally, larger and less elongated ellipse sizes are beneficial for the sensitivity of the COMPASS method.
Consequently, the largest ellipse with values of $\delta_\textnormal{max} = 20^\circ$ and $\delta_\textnormal{min} = 15^\circ$ for the semi-major and semi-minor axes, respectively, yields the lowest isotropic chance probability of $p_\textnormal{val} = 2.8 \times 10^{-3}$.
This result is significantly better than the previously considered ellipse geometry of $(\delta_\textnormal{max},\, \delta_\textnormal{min}) = (10^\circ, 5^\circ)$, which exhibits a chance probability of $p_\textnormal{val} = 1.2 \times 10^{-2}$ in the same benchmark scenario.
However, the specific behavior of the sensitivity for the various ellipse geometries may be characteristic for the simulation setup.
Since the angular scales of existing structures are unknown for an application to data, it is suggested to scan the ellipse geometry in a reasonable range.

\section{Conclusion}

In this work we investigated a novel approach to search for structures in the arrival directions of UHECRs induced by cosmic magnetic fields.
A dynamic vector field tangential to the local celestial sphere is utilized to fit the orientation of elongated patterns.
Thus, elliptically shaped density functions are aligned by the vector field and evaluated in a likelihood ratio with a circular reference model.
This work demonstrates that the orientation of the directional deflections of the GMF is detectable by faint signatures of simulated UHECR sources.
The sensitivity of the method was investigated by means of an astrophysical simulation of uniformly distributed sources where UHECR nuclei are attenuated during propagation in the extragalactic universe.
It was shown that the hypothesis of isotropically distributed arrival directions can be excluded with more than $4 \, \sigma$ Gaussian significance if a maximum spatial density of UHECR sources of $\rho_S = 10^{-2}$~Mpc$^{-3}$ is assumed. 

\section*{Acknowledgments}

We  wish to thank very much N.~Langner for fruitful discussions, and J.~Schulte for valuable comments on the manuscript.
This work is supported by the Ministry of Innovation, Science and Research of the State of North Rhine-Westphalia, and by the Federal Ministry of Education and Research (BMBF).

%

\begin{thebibliography}{}
%
%
\bibitem{Han2017} J. Han, Annual Review of Astronomy and Astrophysics \textbf{55}, (2017) 111-157.

\bibitem{Aab2015} A. Aab et al., Phys. Rev. \textbf{D93}, (2016) 122005.

\bibitem{Aab2014a} A. Aab et al., Phys. Rev. \textbf{D90}, (2014) 122006.

\bibitem{Aab2017} A. Aab et al., Phys Rev. \textbf{D96}, (2017) 122003.

\bibitem{Stanev1996} T. Stanev, ApJ \textbf{479}, (1997) 290.

\bibitem{Harari2000} D. Harari, S. Mollerach, E. Roulet, JHEP 2000 \textbf{02}, (2000) 035.

\bibitem{Harari2002} D. Harari, S. Mollerach, E. Roulet, F. Sanchez, JHEP \textbf{03}, (2002) 045.

\bibitem{Golup2009} G. Golup, D. Harari, S. Mollerach, E. Roulet, Astropart. Phys. \textbf{32}, (2009) 269–277.

\bibitem{Giacinti2010} G. Giacinti, M. Kachelriess, D. V. Semikoz, G. Sigl, JCAP \textbf{1008}, (2010) 036.

\bibitem{Golup2011} G. Golup, D. Harari, S. Mollerach, E. Roulet, JCAP \textbf{1107}, (2011) 006.

\bibitem{Giacinti2011} G. Giacinti, M. Kachelriess, D. V. Semikoz, G. Sigl, Astropart. Phys. \textbf{35}, (2011) 192–200.

\bibitem{Abreu2011} P. Abreu et al., Astropart. Phys. \textbf{35}, (2012) 354–361.

\bibitem{Aab2014b} A. Aab et al., Eur. Phys. J. \textbf{C75}, (2015) 269.

\bibitem{Aab2020} A. Aab et al., JCAP \textbf{6}, (2020) 17.

\bibitem{Erdmann2019} M. Erdmann, L. Geiger, D. Schmidt, M. Urban, M. Wirtz, Astropart. Phys. \textbf{108}, (2018) 74.

\bibitem{Tensorflow2015} M. Abadi et al., Software available from tensorflow.org (2015).

\bibitem{Wirtz2019b} M. Wirtz, M. Erdmann, PoS \textbf{358}, (2019) 470, 36th ICRC.

\bibitem{Erdmann2016} M. Erdmann, G. Mu\"uller, M. Urban, M. Wirtz, Astropart. Phys. \textbf{85}, (2016) 54.

\bibitem{Farrar2017} G. R. Farrar, M. S. Sutherland, JCAP \textbf{5}, (2019) 004.

\bibitem{Renteln2013} P. Renteln, Cambridge University Press, 2013.

\bibitem{Jansson2012a} R. Jansson, G. R. Farrar, ApJ \textbf{757}, (2012) 14.

\bibitem{Farrar2015} G. R. Farrar, N. Awal, D. Khurana, M. Sutherland, PoS \textbf{236}, (2015) 560, 36th ICRC.

\bibitem{Barrera1985} R. G. Barrera, G. A. Estevez, J. Giraldo, Eur. J. Phys. \textbf{6}, (1985) 4.

\bibitem{Aab2018a} A. Aab et al., ApJL \textbf{853}, (2018) 2.

\bibitem{Sommers2000} P. Sommers, Astropart. Phys. \textbf{14}, (2001) 271.

\bibitem{Wilks1938} S. S. Wilks, Annals of Mathematical Statistics \textbf{9}, (1938) 1.

\bibitem{Hilton2012} G. Hilton, T. Tieleman, Lecture, University of Toronto, 2012.

\bibitem{Bister2020}  T. Bister et al., Astropart. Phys., (2020), In Press.

\bibitem{Fenu2017} F. Fenu, PoS \textbf{301}, (2017) 486, 35th ICRC.

\bibitem{Bretz2014} H.-P. Bretz et al., Astropart. Phys. \textbf{54C}, (2014) 110.

\bibitem{Aab2016} A. Aab et al., JCAP \textbf{04}, (2017) 038.

\bibitem{Fisher1953} R. A. Fisher, Proc. R. Soc. A \textbf{217}, (1953) 1130.

\bibitem{Caccianiga2019} L. Caccianiga, PoS \textbf{358}, (2019) 206, 36th ICRC.

\bibitem{Wirtz2020} M. Wirtz, PhD thesis, RWTH Aachen University (2020).

\bibitem{Pshirkov2011} M. S. Pshirkov, P. G. Tinyakov, P. P. Kronberg, K. J. Newton-McGee, ApJ \textbf{738}, (2011) 192.

\end{thebibliography}
%

\end{document}